\documentclass[fleqn,usenatbib]{mnras}

\usepackage{newtxtext,newtxmath}

\usepackage[T1]{fontenc}

\DeclareRobustCommand{\VAN}[3]{#2}
\let\VANthebibliography\thebibliography
\def\thebibliography{\DeclareRobustCommand{\VAN}[3]{##3}\VANthebibliography}

\usepackage{graphicx}	
\usepackage{amsmath}	

\title[Velocity profiles around voids]{Velocity profiles of matter and biased tracers around voids}

\author[E. Massara et al.]{Elena Massara$^{1,2}$\thanks{E-mail: elena.massara.cosmo@gmail.com}, Will J. Percival$^{1,2,3}$, Neal Dalal$^{3}$, Seshadri Nadathur$^{4}$, Sla{\dj}ana Radinovi\'c$^{5}$, \newauthor{Hans A. Winther$^{5}$, Alex Woodfinden$^{1,2}$}
\\
$^{1}$Waterloo Centre for Astrophysics, University of Waterloo, 200 University Ave W, Waterloo, ON N2L 3G1, Canada\\
$^{2}$Department of Physics and Astronomy, University of Waterloo, 200 University Ave W, Waterloo, ON N2L 3G1, Canada\\
$^{3}$Perimeter Institute for Theoretical Physics, 31 Caroline St North, Waterloo, ON N2L 2Y5, Canada\\
$^{4}$Institute of Cosmology and Gravitation, University of Portsmouth, Burnaby Road, Portsmouth, PO1 3FX, United Kingdom\\
$^{5}$Institute of Theoretical Astrophysics, University of Oslo, P.O. Box 1029 Blindern, N-0315 Oslo, Norway
}

\date{Accepted XXX. Received YYY; in original form ZZZ}

\pubyear{2022}

\begin{document}
\label{firstpage}
\pagerange{\pageref{firstpage}--\pageref{lastpage}}
\maketitle

\begin{abstract}
The velocity profile of galaxies around voids is a key ingredient for redshift space distortion (RSD) measurements made using the void-galaxy correlation function. In this paper we use simulations to test whether the velocity profile of the tracers used to find the voids matches the velocity profile of the dark matter around these voids. A mismatch is expected and found in the inner part of voids, where tracers are very sparse. We discuss how this difference is caused by a selection effect where the void centre positions are correlated to the particular realization of the sparse tracers and their spatial distribution. In turn, this then affects the RSD void-galaxy correlation analysis. We show this by evaluating the Jacobian of the real to redshift space mapping using the tracer or matter velocity profile. Differences of the order of 20\% in the velocity profile translate into differences of the order of few percent in the Jacobian. This small discrepancy propagates to the monopole and quadrupole of the void-tracer correlation function, producing modifications of comparable magnitude to those from changes in $f\sigma_8$ at the level of the statistical uncertainties from current analyses. 
\end{abstract}

\begin{keywords}
Cosmology -- large scale structure of universe -- theory
\end{keywords}



\section{Introduction}

\label{sec:intro}

Spectroscopic galaxy redshift surveys map the Universe by measuring the redshifts of a set of galaxies, from which distances can be inferred (e.g. \citealt{Alam_2021}). Galaxy peculiar velocities along the line-of-sight mean that, when mapping redshifts into distances, we obtain a map of the Universe in redshift space, where the position of each object is shifted from its true position by its peculiar velocity \citep{Kaiser_1987}. Therefore, any statistics of the galaxy field measured in redshift space will be affected by peculiar velocities. These velocities contain a lot of cosmological information, including the rate of structure growth, but they also need to be properly modeled not to become a systematic problem for a cosmological analysis. 

Cosmic voids have become a complementary probe to standard analysis techniques, such as Baryon Acoustic Oscillations (BAO) and galaxy-galaxy clustering \citep{Lavaux_2012,Nadathur_2020_lss,2019_whitepaper}. The most popular void-related observable is the void-galaxy correlation function, where void centres are cross-correlated with the galaxy population. If we only consider this statistic, then void identification reduces to identifying a population of void centres. Redshift space and Alcock-Paczynski (AP; \citealt{Alcock_1979}) measurements using the void-galaxy correlation function have been made using SDSS data \citep{SDSS} and have been shown to provide competitive constraints on the growth of structure and the expansion history of the Universe \citep{Hamaus_2016,Hamaus_2017a,Nadathur_2019c,Achitouv_2019,Hawken_2020,Aubert_2020,Hamaus_2020,Nadathur_2020,Woodfinden_2022}. 

Two approaches have been used to locate voids centres when making RSD measurements around them: voids can be identified in redshift space \citep{Hamaus_2016}, or RSD can be approximately removed before voids are found, in a process called RSD reconstruction \citep{Nadathur_2019b,Nadathur_2019c}; in both methods, void centres are then cross-correlated with the redshift space galaxy field. In RSD reconstruction, one find voids from galaxy positions adjusted by subtracting the expected galaxy peculiar velocity from the measured redshift space position, which provides an estimate for the real-space position. The expected peculiar velocity is determined from the gravitational potential estimated from the field of galaxies as is commonly used to enhance the BAO signal \citep{Eisenstein_2007,Burden_2015}. 

For both methods, the mapping between the real and redshift space for the galaxies is usually performed using the velocity profile of galaxies around void centres, which should not be confused with the void-galaxy \textit{pairwise} velocity profile. The velocity profile is the mean radial (spherically averaged) velocity of galaxies around voids, where each velocity is computed relative to the Hubble expansion, while the pairwise velocity profile is related to the difference between the velocity of each galaxy and the velocity of the void. For an individual void, these differ by the void velocity relative to the background Hubble expansion. Averaging over a stack of voids, the velocity profile will be a smoothed version of the pairwise velocity profile, where the smoothing kernel is the distribution of void velocities. The velocity profile $v$ as a function of the distance $r$ from the void centre is
\begin{equation}
    v(r) = \frac{\Sigma_{i=1}^{N_v} \Sigma_{j=1}^{N_p} \, I_r(r_{ij})\, \vec{v}_j \cdot \hat{r}_{ij}}  {\Sigma_{i=1}^{N_v} \Sigma_{j=1}^{N_p} \,  I_r(r_{ij})}\, ,
    \label{eq:vel}
\end{equation} where $N_v$ and $N_p$ are the number of voids and objects (galaxies) considered, $I_r(x) = 1$ if $r-dr<x<r+dr$ and is zero otherwise, $\vec{r}_{ij} = \vec{r}_{j}-\vec{r}_{i}$ is the vector connecting the void centre $i$ and the object $j$, and $\vec{v}_j$ is the velocity of the object $j$. This observable is therefore the number of void-object pairs weighted by the object velocity, and normalized by the number of pairs. The mapping between the void-galaxy correlation function $\xi(r)$ in real space, and $\xi(s,\mu)$ in redshift space, is usually described as \citep{Nadathur_2019a}\footnote{See \cite{Woodfinden_2022} for a detailed explanation of why this model is equivalent to the streaming model \citep{Fisher_1995}.}
\begin{equation}
    1+\xi(s,\mu) = \int d\Tilde{v}\, p(\Tilde{v},r)\, \left[ 1+\xi(r) \right]\,J^{-1}(r,\mu,v(r))\, ,
\end{equation}
where $\mu$ is the cosine of the angle between the direction of the void-galaxy pair and the line of sight $\hat{r}_{\rm los}$, $p(\Tilde{v},r)$ is the probability distribution function of incoherent velocities $\Tilde{v}$ once the coherent outflow described by the velocity profile $v(r)$ has been subtracted, and $J(r,\mu)$ is the Jacobian of the transformation $\vec{s} = \vec{r}+\hat{r}_{\rm los}v(r)\mu/aH$, and it depends on the velocity profile only. The simplest model \citep{Cai_2016},
\begin{equation}
    1+\xi(s,\mu) = \left[ 1+\xi(r) \right] \,J^{-1}(r,\mu,v(r))\, ,
    \label{eq:xi_r_to_s}
\end{equation}
considers only the (coherent) galaxy velocity profile to map the cross-correlation between the two spaces.

In the literature, the velocity profile of the \textit{galaxies} is often assumed to match at all scales the velocity profile of the \textit{matter} around the same voids. Even under the assumption that locally galaxies follow the same velocity field as the matter, which we assume to be true, the profiles can differ due to selection effects. The assumption that the profiles match brings the advantage of being able to model the galaxy profile as a matter profile, that could ideally be directly connected to cosmology and predicted from first principles. Practically, it is usually computed using the continuity equation, and in particular a linearized version of it that maps the fully-nonlinear matter enclosed density profile $\Delta(r)$ 
\begin{equation}
    \Delta(r) = \frac{3}{r^3}\int_0^r dx \, x^2\delta(x),
\end{equation}
into an estimation for velocity profile $v(r)$, which is close to (but not exactly matching) the fully-nonlinear velocity profile measured in simulations \citep{Massara_2018}, via 
\begin{equation}
    v(r) = -fHar \,\frac{\Delta(r)}{3}\, ,
    \label{eq:d_to_v}
\end{equation}
where $f$ is the linear growth rate, $H$ is the Hubble parameter, $a$ is the scale factor, $r$ is the distance from the void centre and $\delta(r)$ is the density profile inside a shell at distance $r$. 

In this paper we consider the possibility that the velocity profile of the galaxies and of the matter are not equal at all scales. Galaxies are biased and sparse tracers of the matter field. Although their sparsity becomes particularly relevant in regions devoid of matter, we have to use galaxies to identify voids. Finding voids in a sparse sample can make their properties very sensitive to the shot-noise of the sample. In particular, the void centre definition might be highly affected by it, since it correlates with the particular Poisson realization of galaxies used to identify voids. If the void centre is shifted with respect to the minimum in the matter density field, the distribution of matter around the void centre can be asymmetric and differ from the distribution of galaxies around it. This can bring two main effects: the matter and velocity profiles around the void centre are different, and the ratio between the galaxy and matter density profiles is different from the galaxy bias.  

We use simulations to explore the effect of the sparsity of the tracers used to find void centres on the recovered velocity and density profiles around voids. We will build toy models to describe the effects, and then measure them in {\it N}-body simulations, where halos are used as biased and sparse tracers to identify voids. We will implement different void finders and describe how sparsity affects each. 

We use a subset of the {\it N}-body simulations in the Quijote suite \citep{Quijote_2020}. The selected simulations are $50$ realizations with $512^3$ dark matter particles in a volume equal to $1 h^{-3}$Gpc$^3$ and a flat $\Lambda$CDM cosmology with parameters $\Omega_m = 0.3175$, $\Omega_b=0.049$, $h=0.6711$, $n_s=0.9624$, $\sigma_8=0.834$ and $M_\nu = 0$. These simulations have been run using initial conditions generated at redshift $z=127$ via second order perturbation theory (2LPT) and with gravitational softening length set to $1/40$ of the mean interparticle separation. Haloes have been identified in the matter field using the Friends-of-Friends (FoF) algorithm \citep{Davis_1985} with linking length parameter $b = 0.2$. The halo catalogs contain halos with at least $20$ particles, thus the minimum halo mass is $1.31\cdot 10^{13}h^{-1}$M$_{\odot}$. In our analysis, we consider cold dark matter snapshots and halo catalogs obtained at redshift $z=0$, and we identify voids in both the matter and halo field.

\section{Void finders} \label{sec:finders}

In this analysis we use four different void finders: a void finder that identifies spherical regions with density contrast below a certain threshold, and three watershed void finders that do not assume any particular shape for the voids. Among these, two rely on the same algorithm to identify the regions that are voids, but assume a different definition for the void centres. Below we describe the details of each void finder and highlight their differences.   

\subsection{Spherical}
We use the spherical void finder implemented in the \textsc{Pylians}\footnote{\url{https://github.com/franciscovillaescusa/Pylians3}} library and designed to identify non-overlapping voids as spherical regions of different sizes with overdensity $\Delta=\rho/\bar{\rho}-1$ below a given threshold. 
The algorithm takes as input the overdensity threshold and the list of void radii to consider, and it works as follows. First, it assigns the tracers to a mesh and computes the density field in each voxel. Then it smooths the density field with a top-hat filter of scale $R_{\rm v}$ corresponding to the largest void radius given as input. Voxels whose smoothed density is below the declared threshold are sorted from low to high values. The algorithm considers the voxel with the lowest smoothed density and identifies as a first void the region centred on that voxel, containing all the cells within a distance equal to $R_{\rm v}$. Subsequently, it considers the second most underdense voxel in the smoothed field, and all the cells within $R_{\rm v}$ from it. If none of these cells belong to a previously identified void, then they are assigned to the new one and a new void is found; otherwise, the candidate is discarded. 
The algorithm continues inspecting all the selected voxels before switching to the next largest void size given in input. At that point, the smoothed density field is recomputed using the updated value for $R_{\rm v}$ and voxels with smoothed overdensity below the thresholds are sorted and inspected. This procedure is repeated for all the void sizes given as input, and for all the selected voxels with smoothed density below threshold. The void centre is defined as the location where the smoothed field is below the threshold. Since the field is smoothed on large scales (at the void size $R_{\rm m}$), this centre definition is non-local.

Starting the void identification using large smoothing scales and lower overdensities at fixed $R_{\rm v}$ allows us to identify larger voids first, in order to avoid the void-in-void problem of double-counting small voids that are part of larger ones. In this work we report on the analysis using voids defined with threshold  $\Delta_v = -0.5$, but we have tested additional values for the threshold ($\Delta_v = -0.7,-0.8$), which all provide similar results. We consider a mesh with $768^3$ voxels, and we bin the void sizes in 16 multiples of the voxel size, $R_{\rm v}=\left[ 13 - 73\right]\cdot l_{\rm voxel}$, where $l_{\rm voxel}\simeq 1.3h^{-1}$Mpc.
The void radii are set by the choice of the smoothing scales. The finder identifies all voids of intermediate sizes between two smoothing scales and assigns them a radius equal to the smaller smoothing scale. Let’s call $dR_{\rm v}$ the difference between adjacent smoothing scales (or void radii); if the size of a void is defined as the scale at which the enclosed density is equal to the threshold, then voids with size between $R_{\rm v}$ and $R_{\rm v}$+$dR_{\rm v}$ are assigned to the radius $R_{\rm v}$ because 1) we start identifying voids using the largest smoothing scale and decrease it with time, 2) the finder requires that the enclosed density must be {\it below} the threshold, and 3) the density increases with distance from the void radius.

\subsection{ZOBOV}
\texttt{ZOBOV} \citep{zobov} is a watershed algorithm that identifies voids by looking at the topology of the large-scale structure. First, it divides the space using a Voronoi tessellation: a Voronoi cell is assigned to each tracer and its extension is defined as the space closer to that tracer than to any other in the catalog. The tessellation naturally gives an estimation of the local density as the inverse of volume of each Voronoi cell; indeed, larger cells belong to tracers that are more isolated. \texttt{ZOBOV} uses this estimation of the local density to identify local minima in the density field. The second step identifies basins as groups of Voronoi cells via a watershed procedure. The algorithm starts from the deepest local minimum and adds adjacent Voronoi cells whose density is larger than previously added ones. If the next adjacent cell is less dense, then it has found a saddle point and it stops adding cells along that direction. When all possible cells are added to the basin of the deepest local minimum, it starts adding cells to the next deepest minimum to identify a second basin. Once all the space is divided into basins, an optional final step can be taken to merge neighbouring basins together into the final voids. The criterion to merge basins usually depends of ratios between the density of the Voronoi cell that is a local minimum and the cells at the boundaries of the basins. Given a set of \texttt{ZOBOV} voids, we now consider two ways of defining the void centre.

\subsubsection{VIDE} \label{subsec:VIDE}

\texttt{VIDE}\footnote{\url{https://bitbucket.org/cosmicvoids/vide_public/wiki/Home}} \citep{Sutter_2015} is a wrapper around \texttt{ZOBOV} that can identify voids in both periodic boxes and survey data, and it provides tools to compute different void-related statistics. We will refer to VIDE voids as voids identified using \texttt{VIDE}, where each basin is a final void, i.e. we do not merge basins in agreement with the latest papers using \texttt{VIDE}. In accordance with these papers, VIDE voids will have centres defined as the volume-weighted barycentre
\begin{equation}
    \vec{x}_v = \frac{\Sigma_i \vec{x}_i\textit{V}_i}{\Sigma_i \textit{V}_i},
\end{equation}
where $\vec{x}_i$ is the position of the $i$-th Voronoi cell belonging to the void, and $\textit{V}_i$ is its volume (normalised in units of the mean volume of all cells). This definition for the void centre takes into account the position of all of the tracers belonging to the void, and is therefore non-local. Most of the tracers are in the outskirts of voids, but their importance in the determination of the void centre location is down-weighted by their small Voronoi volume. A size is associated to each void corresponding to the radius of a sphere with the same volume as that obtained by summing the volumes of the Voronoi cells belonging to it. 

\subsubsection{ZOBOV-circumcentre} 

ZOBOV-circumcentre voids are identified using the wrapper \texttt{Revolver}\footnote{\url{https://github.com/seshnadathur/Revolver}} around \texttt{ZOBOV} and, as with VIDE voids, they consist of basins that are not merged. ZOBOV-circumcentre and VIDE catalogs will share the same number of voids with the same sizes, but voids have different centres. For ZOBOV-circumcentre, the location of the void centre is computed as the point of lowest density within the void, identified via the following procedure. In each void, the largest Voronoi cell -- corresponding to the lowest density -- is identified, together with the three next-largest Voronoi cells that are mutually adjacent to it. The locations of the tracers associated with each of these Voronoi cells define a tetrahedron, and the circumcentre of this tetrahedron is taken as the centre of the void \citep{Nadathur:2015a}. This construction ensures that the void centre corresponds to the centre of the largest empty sphere that can be inscribed the void.
Its location is only indirectly informed by the location of all the tracers in each void, and mainly depends on the position of the four most isolated ones. Thus, it is a local definition. 

\subsection{Voxel}

Voxel voids are watershed voids that do not use Voronoi tessellation: the density field is evaluated on a mesh. The algorithm to identify this type of voids is coded in \texttt{Revolver}. The void finder places the tracers on a mesh whose size $N_{\rm mesh}$ is determined by the mean number density of the tracers $n_t$: $N_{\rm mesh} = L_{\rm box} \, 2 \, (4 \pi \, n_t/3)^{1/3}$, where $L_{\rm box}$ is the size of the N-body simulation. The density field is then smoothed using a Gaussian filter at size equal to $n_t^{-1/3}$ and local minima are identified in the density field over the voxels. Basins are generated around each local minimum in the same way as in \texttt{ZOBOV}: adjacent voxels with increasing overdensity are added to the basin and this process stops when the next voxel is more empty than the previous one. Each final basin corresponds to a Voxel void. 

As for \texttt{ZOBOV}, these voids can have arbitrary shapes and their size is computed as the radius of a sphere containing the same volume as the sum of the individual voxels belonging to the void. The default void centre is the location of the voxel that is a local minimum of the density, and it is therefore a locally defined centre. Additionally, we consider an inverse-density-weighted barycentre defined as the geometrical centre of the tracers weighted by the inverse of the density of the voxel they belong to. This second definition is very similar to the centre determination in VIDE voids, and it is a non-local definition.

\section{Toy Model}
\label{sec:toy_model}
We build toy models to understand the effect of sparsity on the relevant properties observed for a particular void. 

\begin{figure*}
\includegraphics[width=0.35\textwidth]{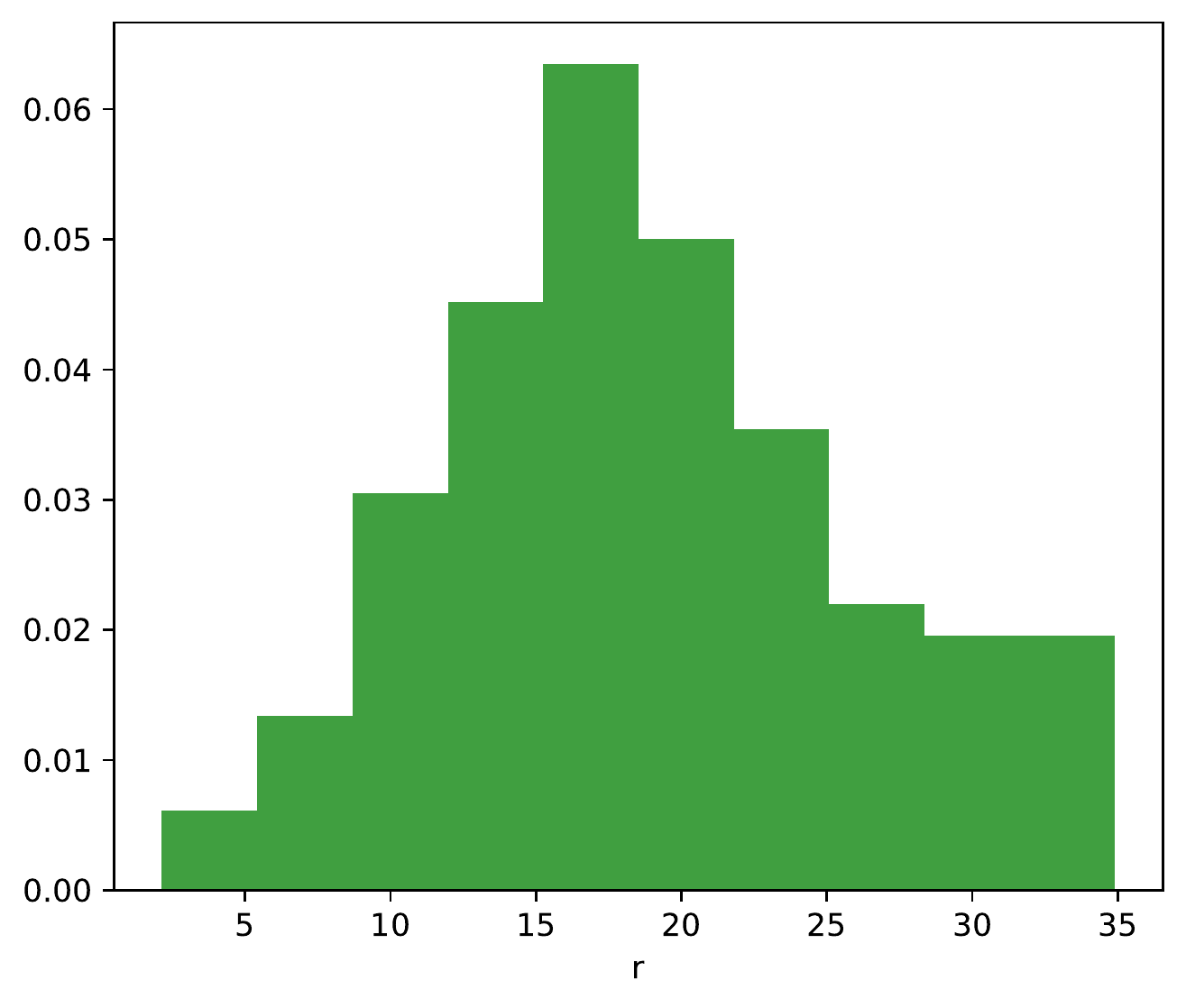}
\includegraphics[width=0.35\textwidth]{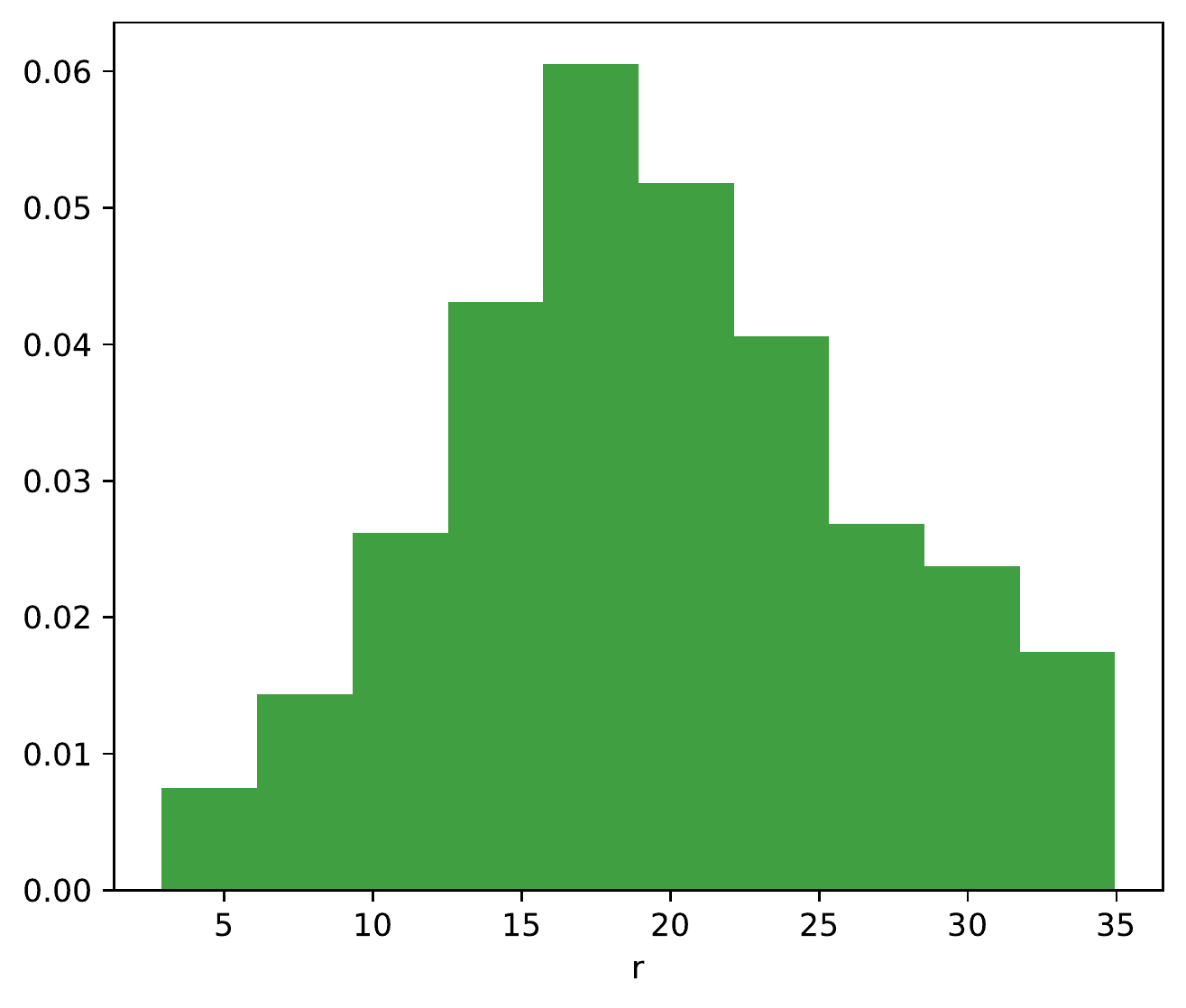}
\caption{Distribution of shifts between void centres in the original distribution and the subgroup particles. The initial particles are assumed to follow Voxel (left) and ZOBOV circumcentre (right) void-like density and velocity profiles. The centre in the subgroup particles has been identified as the Voronoi cell with larger volume.   \label{fig:pdf_shift}}
\end{figure*}

\subsection{Test for a void centre shift}
\label{sec:shift}

We first test the likelihood of an offset in void centre. 
That is, we want to understand if a void finder applied to a sparse subsample of a matter field can identify a void centre position that is shifted with respect to the analogous void centre identified in the full matter field. Since we have in mind halos or galaxies as sparse tracers, we will subsample an initial matter field to reproduce the average number of halos in the simulations used later on. 

To simulate a matter field around a void, we create a 3D distribution of matter particles that is spherically symmetric around the origin of the coordinates, has a density profile that reproduces the one measured around voids in the simulations, and has a mean background density that matches the one of cold dark matter in the same simulations. 
Notice that this setup is an idealization since voids are not isotropic individually. 
We randomly select $0.3\%$ of the particles, in order to mimic a set of halos and to match their mean number density in simulations. We apply a Voronoi tessellation to this subgroup, and identify the void centre as the location of the Voronoi cell with largest volume. Indeed, if a cell has a large Voronoi volume it means that it is isolated, and thus in an low density region. 
This definition for the void centre does not exactly match any of the void finders considered in this paper, but it is assumed for simplicity, and because it is a local definition of the centre similar to the one of ZOBOV-circumcentre and Voxel voids (see Section~\ref{sec:finders} for further details on these void finders). We calculate the distance of the largest Voronoi cell from the origin, which is the true centre of the void in the full sample of matter particles.  Figure~\ref{fig:pdf_shift} displays the distribution of shifts between the two void centres after computing 500 realizations of the setup explained above. The left panel shows the results when assuming a Voxel-like  density profile for the full particle sample, while the right panel considers a ZOBOV-circumcentre profile. In both cases, the assumed void has radius $R_{\rm v} \sim 30h^{-1}$Mpc. As expected, the void centres in the subgroup have shifted from the location of the centre in the full sample. The distribution of these shifts peaks around $15-20h^{-1}$Mpc.

\begin{figure}
\includegraphics[width=0.9\columnwidth]{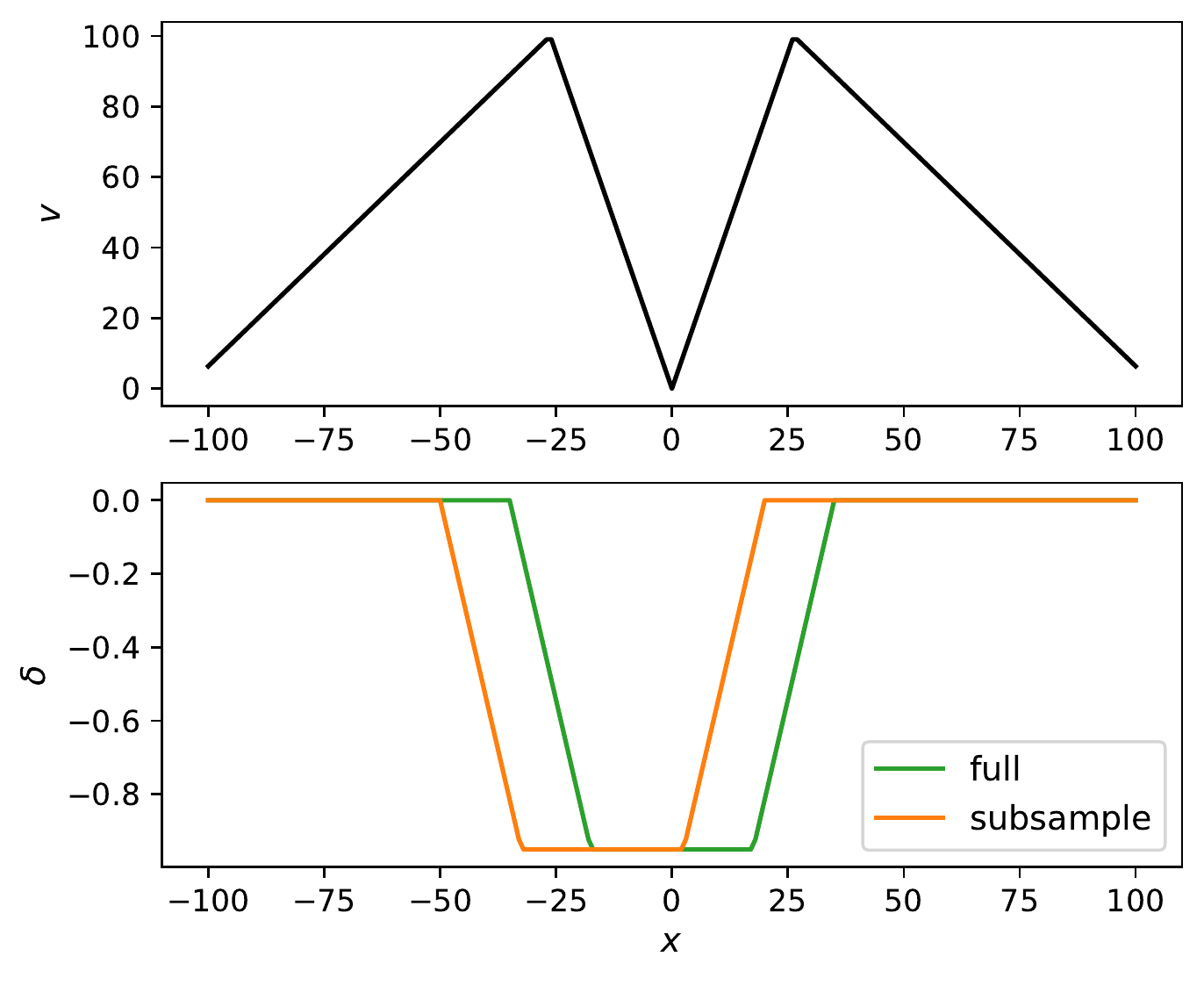}
\caption{Values of the radial velocity (top) and density (bottom) of particles in the 1D toy model. The green line shows the distribution of the full matter field centred at $x=0$, and the orange line displays the distribution of subsampled matter field centred at $x=-15$.
\label{fig:toy_model_1D_x} }
\end{figure} 
\begin{figure}
\includegraphics[width=0.9\columnwidth]{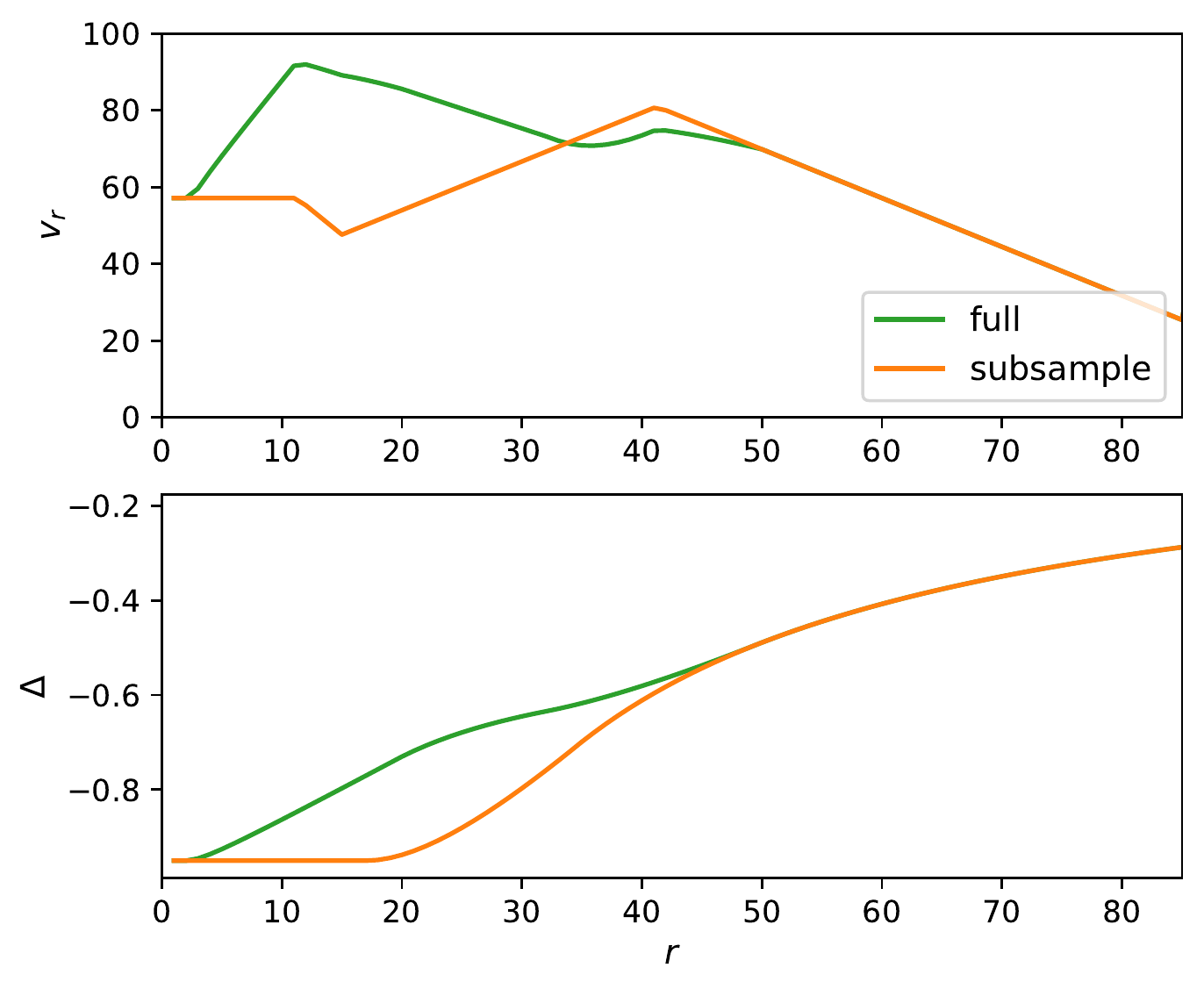}
\caption{Velocity (top) and enclosed density (bottom) profiles in the 1D toy model. Green/orange lines show the profiles of the full/subsampled matter field. All profiles are computed with respect to the centre of the subsampled distribution $x=-15$. 
\label{fig:toy_model_1D_r} }
\end{figure} 

\subsection{The consequences of a void centre shift in 1D}

\begin{figure*}
\centering
\includegraphics[width=0.665\columnwidth]{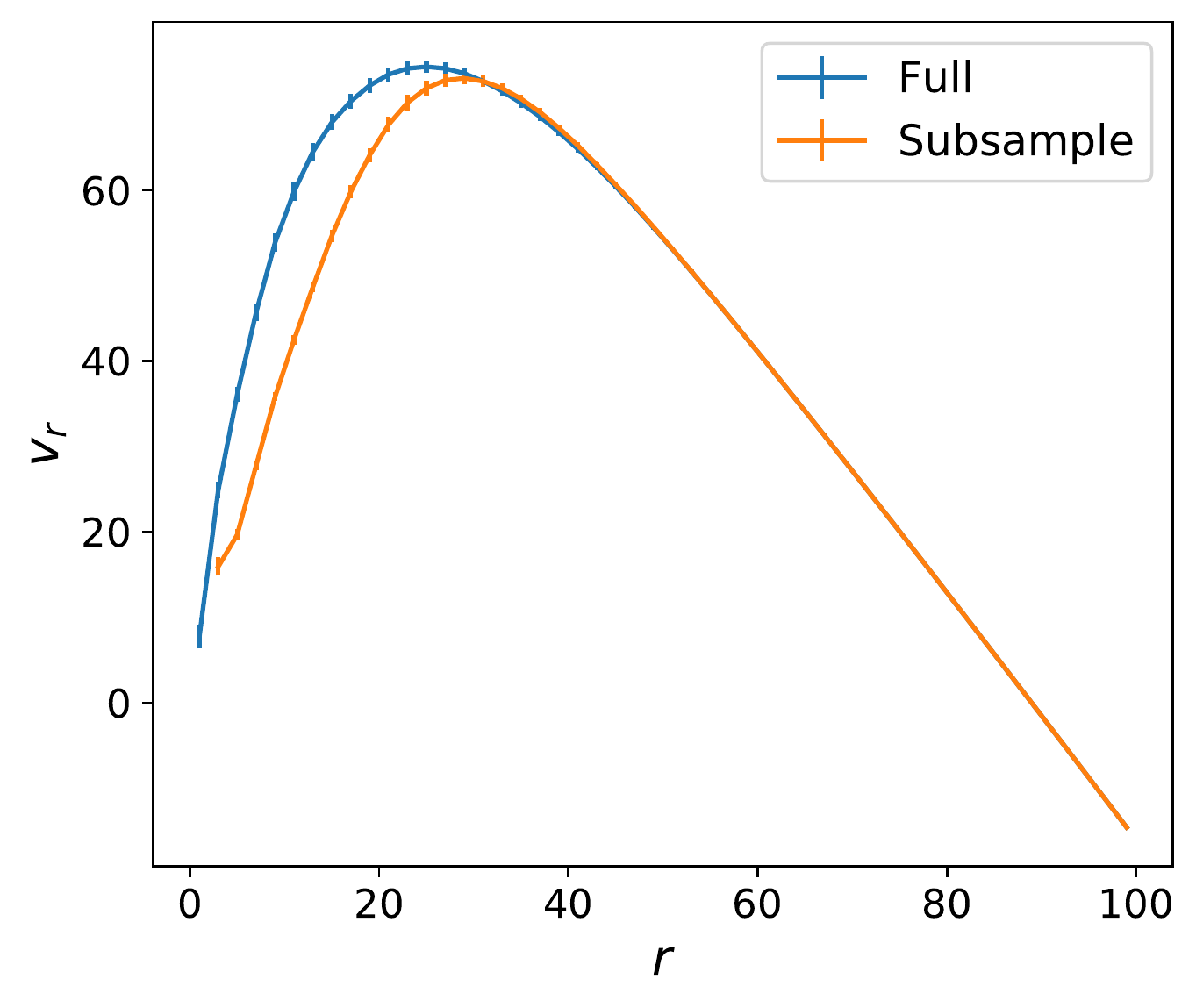}
\includegraphics[width=0.69\columnwidth]{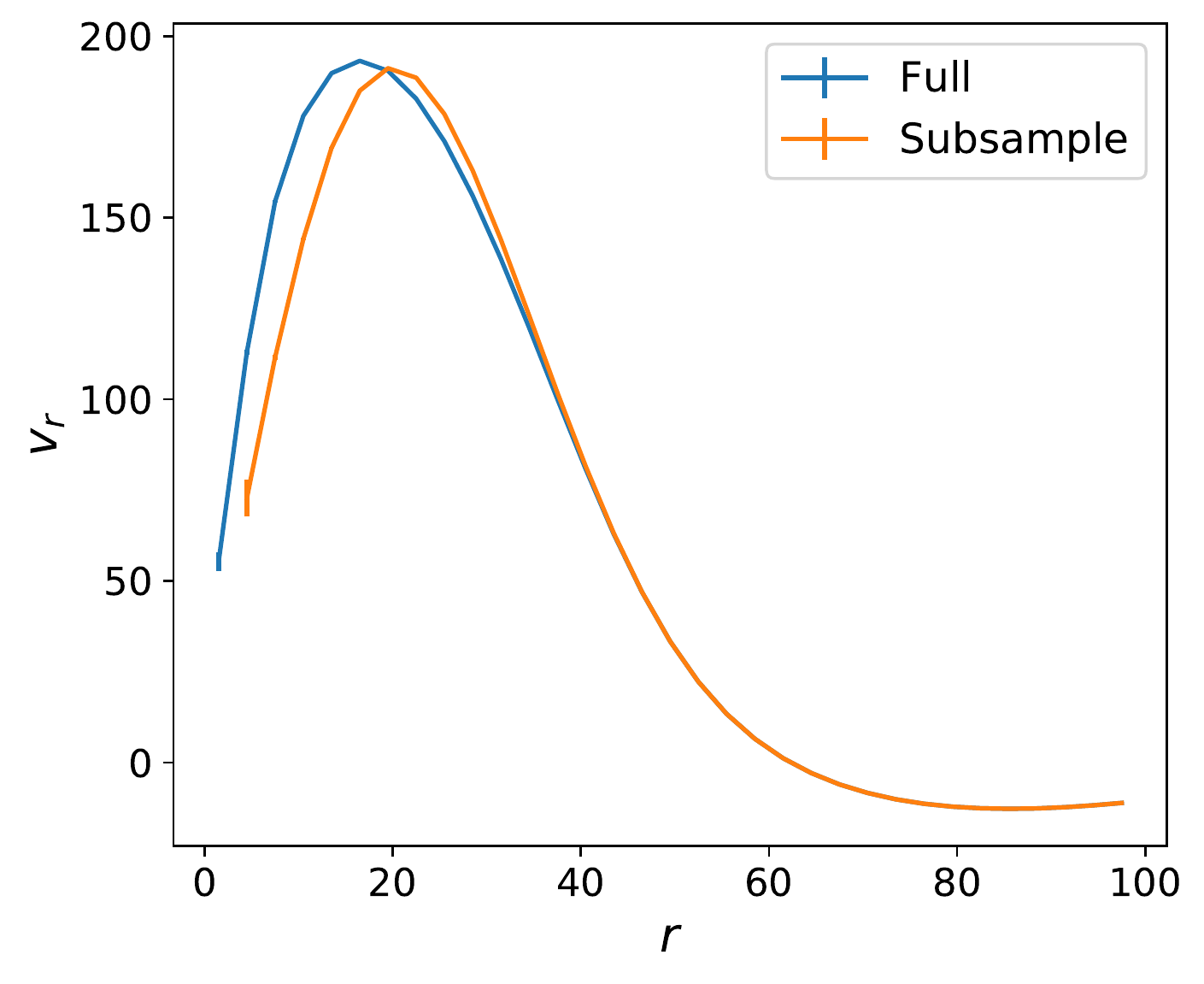}
\includegraphics[width=0.69\columnwidth]{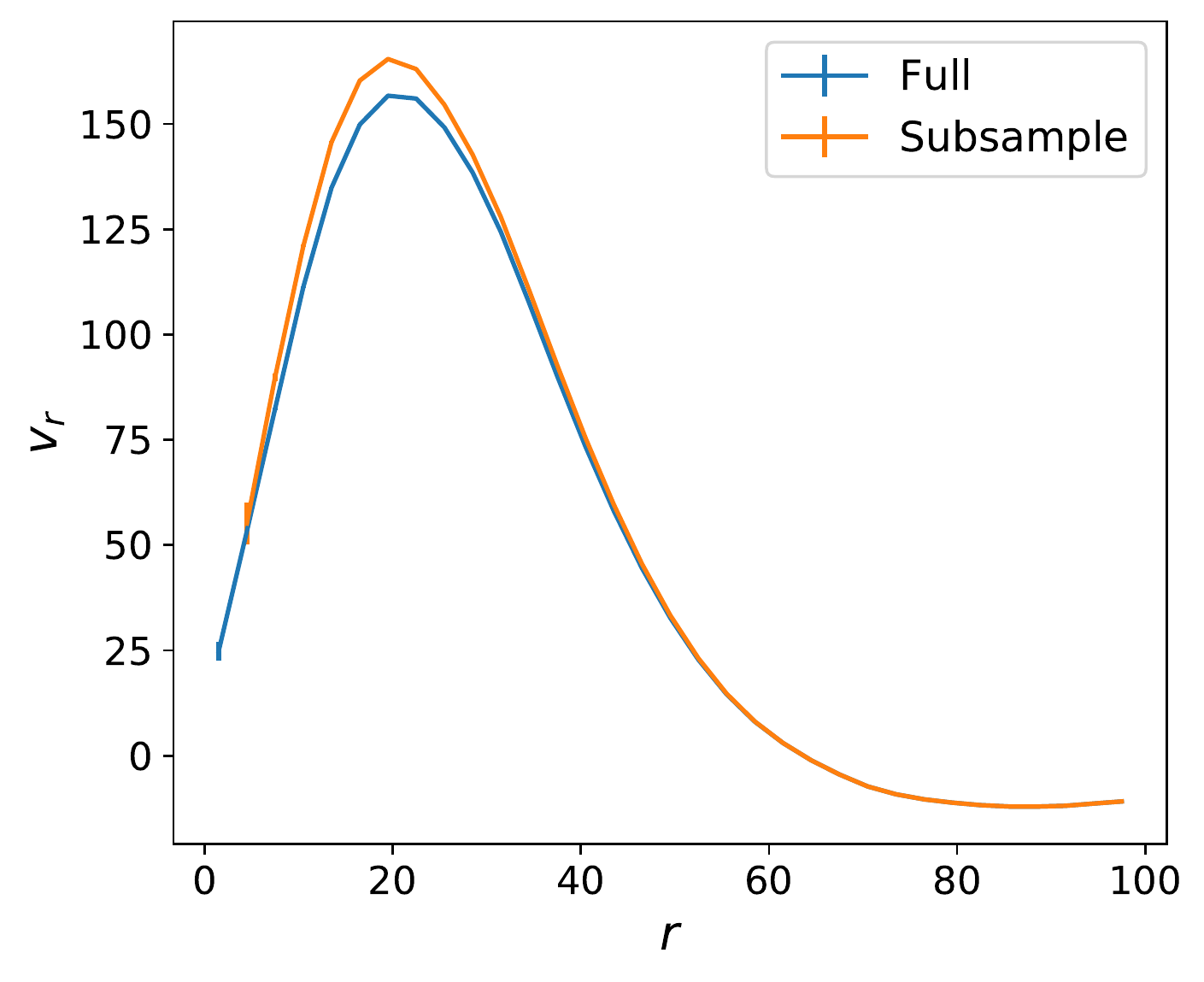}
\caption{Velocity profiles of the full (blue) and subsampled (orange) matter field around void centres in the subsample in 3D toy models. The left panel (a) displays the case of idealized voids where the spatial distribution of particles and their radial velocity follow Equations~\ref{eq:1d_matter} and~\ref{eq:1d_velocity} with $R_{\rm v}=30$ and Gaussian distribution for shifted centres $p(\mu=15,\sigma=7)$. The central (b) and right (c) panels have been obtained from the case where particles follow the density and velocity profiles measured in simulations around Voxel voids with radius $R_{\rm v} = 30 h^{-1}$Mpc, and the offsets for the void centres follow a Gaussian distribution $p(\mu=10,\sigma=5)$ and $p(\mu=20,\sigma=10)$, respectively. Each line displays the mean of 30 realizations of the considered toy model, and errors have been computed as the standard deviation divided by $\sqrt{30}$.} \label{fig:toy_model_3D}
\end{figure*}

We now assume that the void centres identified in a sparse tracer field are shifted with respect to void centres in the matter field (full sample) and investigate its consequences. First, we consider a one dimensional (1D) toy model where the minimum of the matter overdensity is at the centre of coordinates. The toy model is built as follows. We consider a distribution of ``matter" particles that resembles a void-like density profile,
\begin{equation} \delta = \begin{cases}
    -0.95 & |r| < R_{\rm v}/2 \\
    1.9( |r|/R_{\rm v}-1) & R_{\rm v}/2< |r| <R_{\rm v} \\
    0 & |r| > R_{\rm v}
    \end{cases}
    \label{eq:1d_matter}
\end{equation}
where $r$ is the distance from the void centre and $R_{\rm v}$ is the void size. This function is a simplified version of the distribution of matter around voids containing the main features of a real one: the void is almost empty in the very inner part up to half the void radius, while at larger distances from the void centre the density grows until it reaches the mean background density at the void radius. We give each particle a radial velocity $v$ that depends on its distance $r$ from  the origin of coordinates
\begin{equation} v = \begin{cases}
    4/3 \, C\,|r|/R_{\rm v}    & |r| < 3R_{\rm v}/4 \\
    4/3 \, C \left[ 1-|r|/(3R_{\rm v})\right]   & |r| > 3R_{\rm v}/4
    \end{cases}
    \label{eq:1d_velocity}
\end{equation}
where $C$ is the maximum value of the velocity reached at a distance $r=3R_{\rm v}/4$, corresponding to an inflexion point of a realistic density profiles; on larger and smaller distances the velocity decreases, reproducing qualitatively the velocity profiles measured in simulations. 

An example of this setup for an idealized void of size $R_{\rm v} = 30 h^{-1}$Mpc and maximum velocity $C=100$ is shown in Figure~\ref{fig:toy_model_1D_x}. The top panel displays the particles' velocity, which is zero at the matter void centre (origin of the coordinates), increases moving away from the centre up to distances equal to the void size, and starts decreasing to larger distances. The bottom panel shows the two distributions in the toy model. The distribution of matter particles is described by the green line: it follows Equation~\ref{eq:1d_matter} and it is centred at $x=0$. 

We consider a subgroup of matter particles that corresponds to Equation~\ref{eq:1d_matter} centred at $x=-15$ (the distribution is illustrated by the orange line in Figure~\ref{fig:toy_model_1D_x}). The spatial distribution of particles in the subgroup is shifted with respect to the full sample, implying that also the void centre identified using the subgroup is shifted compared to the one in the full matter field.
Notice that, for simplicity, we assumed the two distributions to be equal (but centred around a different minimum: $x=0$ for the green line and $x=-15$ for the orange line), however this assumption might not be valid in general. The two distributions could be different and be centred around distinct locations, or they could be equal and centred around the same location. In this last case, also the velocity profile of the two samples would be equal. The velocities of the subgroup will depend on the distance of each particle from the origin of the coordinates, as shown in the top panel. This assumption is automatically satisfied for subsamples, but should be valid for tracers (halos and galaxies) as well, since their velocity should depend on the gravitational potential set by the full matter field. 

If we were to identify the void centre using the subgroup, its centre would be shifted with respect to the centre of the trough in the full sample, but it is this point that we would average around to compute velocity and density profiles of the matter and of the subsample. These profiles are presented in Figure~\ref{fig:toy_model_1D_r} and are computed analytically (without the need of generating particles) using Equations.~\ref{eq:1d_matter} and~\ref{eq:1d_velocity}. For the toy model, the distribution of subsampled particles on both sides of this void centre are equal, but this is not true for the full sample (see bottom panel of Figure~\ref{fig:toy_model_1D_x}). This asymmetry makes the enclosed density profiles $\Delta(r) = \int_0^r \delta(x) dx /r$ of the full sample (green) and of the subgroup (orange) differ at small distances $r$ from the void centre, as displayed in the bottom panel of Figure~\ref{fig:toy_model_1D_r}. Moreover, this asymmetry coupled to the particular shape of the velocity --- symmetric with respect to the centre of the void in the full sample --- gives rise to different velocity profiles for the subgroup (orange) and full matter (green) particles, as shown in the top panel of Figure~\ref{fig:toy_model_1D_r}. The difference between these velocity profiles is only due to the relatively shifted density distributions; in this case, the velocity profile of the full sample is larger than that of the subgroup. If the spatial distributions of the full and subsample were equal and centred at the same location, then their velocity profiles would also be equal when computed around the same centre, even if that is the wrong location for the void centre; being at a wrong location would change the amplitude and shape of both velocity profiles in the same way compared to the ones computed with respect to the true void centre, but would not introduce any mismatch between the full sample and the subgroup.

\subsection{The consequences of a void centre shift in 3D}

We can now generalize this toy model to 3 dimensions. We generate a 3D distribution of particles with velocity that depends on their distance from the centre of origin. Particles will be distributed spherically around the origin with density and velocity as in Equations~\ref{eq:1d_matter} and~\ref{eq:1d_velocity} with $R_{\rm v} = 30 h^{-1}$Mpc and $C=100$. Following the logic of the 1D toy model, we select a subgroup of particles such that their distribution is spherically symmetric around a different centre, and such that their density profile around it mimics the density profile of the full sample around the origin. Note that the level of subsampling is not important here, since we are not predicting the offset of the void centers and profiles depend only on over-densities and not on the absolute value of the mean density. We repeat this procedure for 10 shifted centres whose distance from the origin is Gaussian distributed with different values of mean and standard deviation. We then measure the velocity profile of the subgroup and of the full sample around the 10 shifted centres, and repeat this procedure 30 times to obtain an estimation of the errors on the velocity profiles. Results for a Gaussian distribution of center offsets with mean equal to $15$ and standard deviation equal to $7$ (to mimic the distribution of center offsets in Figure~\ref{fig:pdf_shift}) are shown in panel (a) of Figure~\ref{fig:toy_model_3D}. As in the 1D case, the velocity profile of the full sample (blue line) is larger than that of the subgroup (orange line). This difference disappears on larger distances $r$ from the void centre.  

In order to have more realistic results, we re-run the same 3D toy model but assuming a different spatial distribution for both the full sample (around the origin) and for the subgroup (around the shifted centre) that follows a density profile measured in simulations around voids of size $R_{\rm v}=30-35h^{-1}$Mpc. Similarly, the outflow velocity of these particles will vary with their distance from the origin of coordinates to reproduce the velocity profile measured in simulations. We measure the velocity profile of the subgroup and of the full sample around 10 shifted centres, as if we had identified voids using the subgroup. Results for a mean of 30 realizations of 10 voids each are shown in panels (b) and (c) of Figure~\ref{fig:toy_model_3D}, with Gaussian distributions for the centre offsets being $p(\mu=10,\sigma=5)$ and $p(\mu=20,\sigma=10)$, respectively. The particular choice of these distributions has been made to consider mean values above and below the distribution in panel (a) of the same figure, and to understand the consequences of these variations. When the average shift between the void centre in the subgroup and in full sample is small (e.g. equal to $10$ as in panel (b)) the velocity profile of the full sample is larger than that of the full sample at small distances from the void centre. On the contrary, larger shifts (equal to $20$ as in panel (c)) give the opposite behaviour: the velocity profile of the subgroup is larger than that of the full sample at distances close to the void size (here $R_{\rm v}\sim35$). 

To summarize, we have shown that if there are two samples of particles, where one is obtained by subsampling the other, and they present relatively shifted local minima (void centres) and density distributions, then their velocity profiles around the same centre can differ. Moreover, how much they differ will depend on a combination of factors, including the size of the void, the distance between void centres in the two samples, and potentially differences in the density profile of the two samples around their minimum (although we assumed them to be negligible in our toy model). Therefore, if sparsity induces an offset between the density distributions and thus between the void centres in two samples (as shown in Section~\ref{sec:shift}), then it is likely that the velocity profiles of the two samples around the same centre are different. 

\section{Sparsity of tracers in simulations}
\label{sec:subs_m}
\begin{figure*}
\includegraphics[width=0.64\columnwidth]{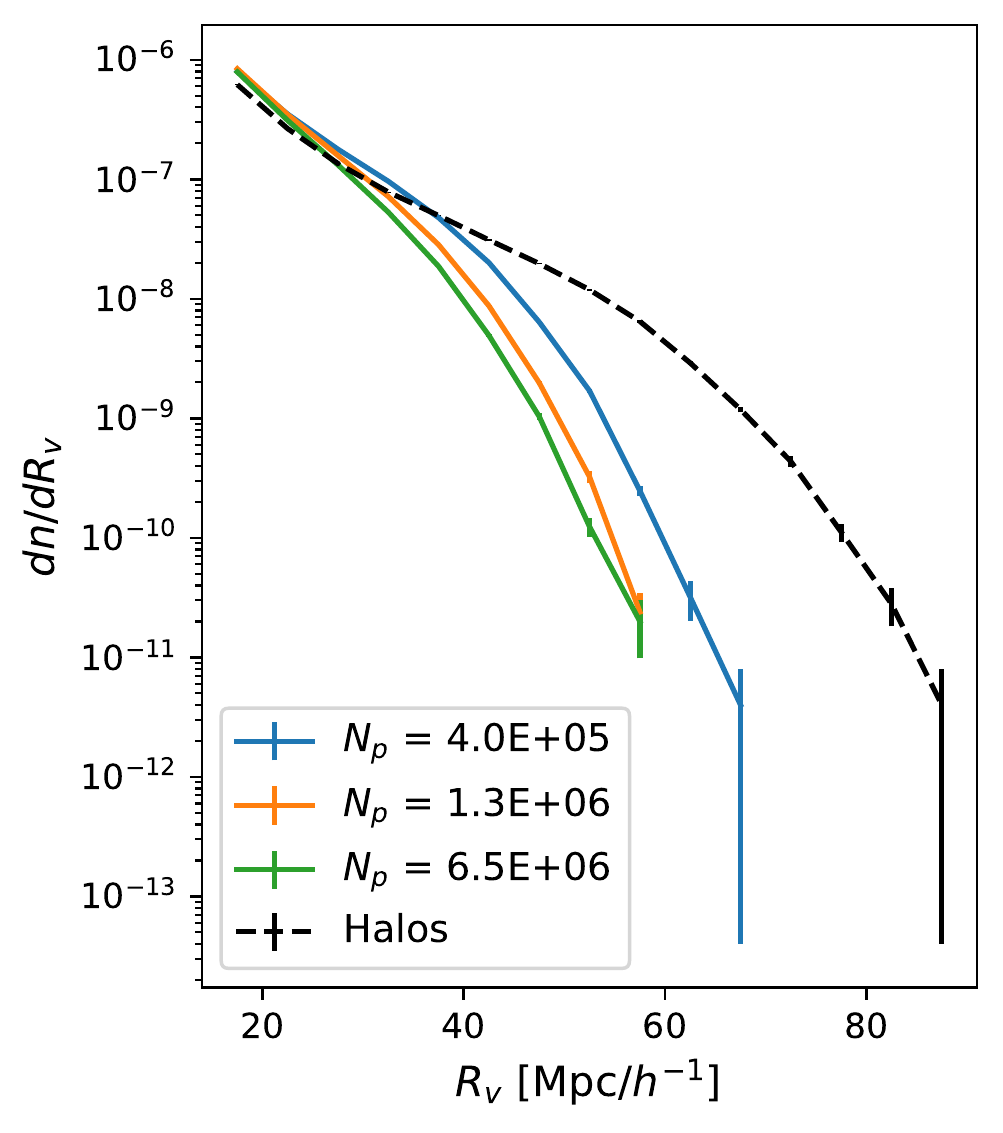}
\includegraphics[width=0.64\columnwidth]{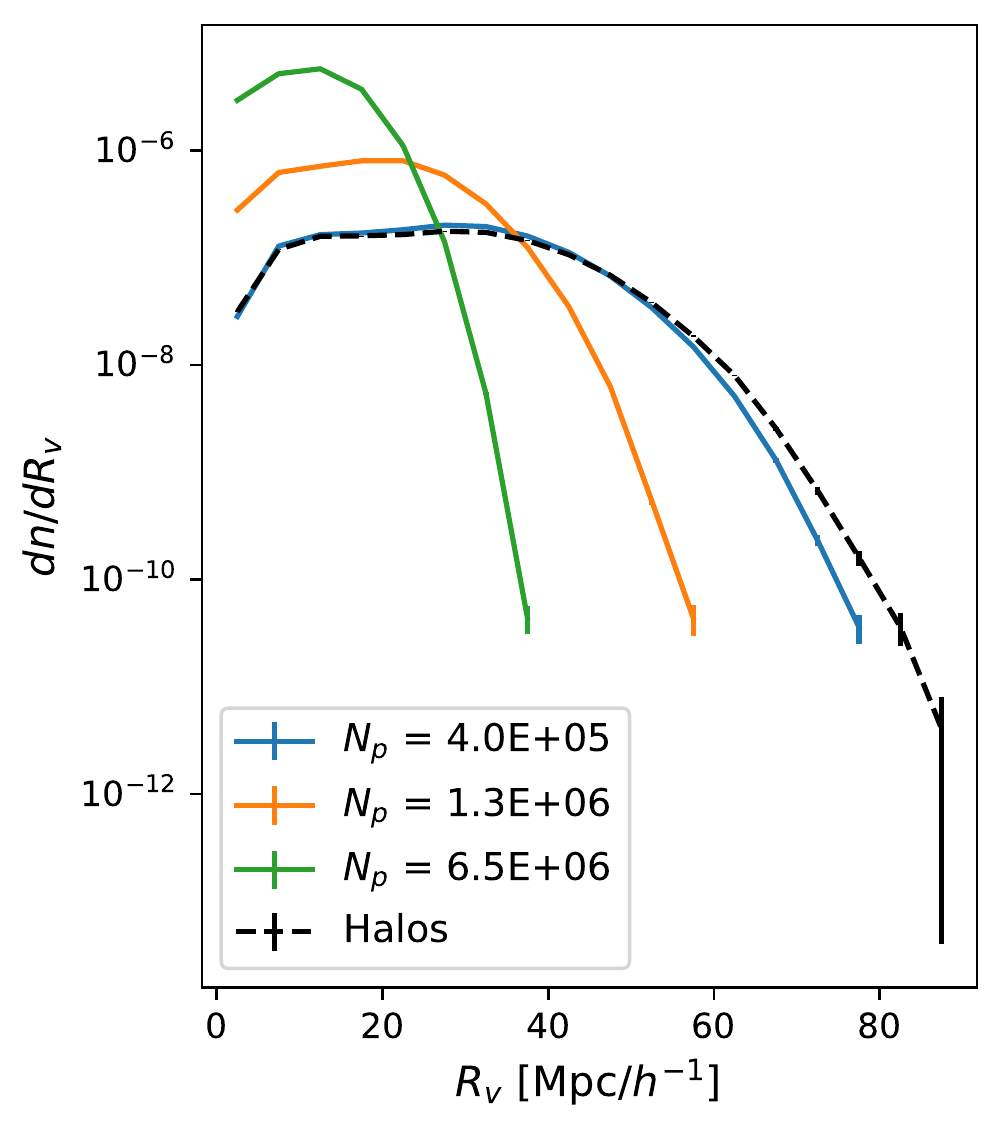}
\includegraphics[width=0.65\columnwidth]{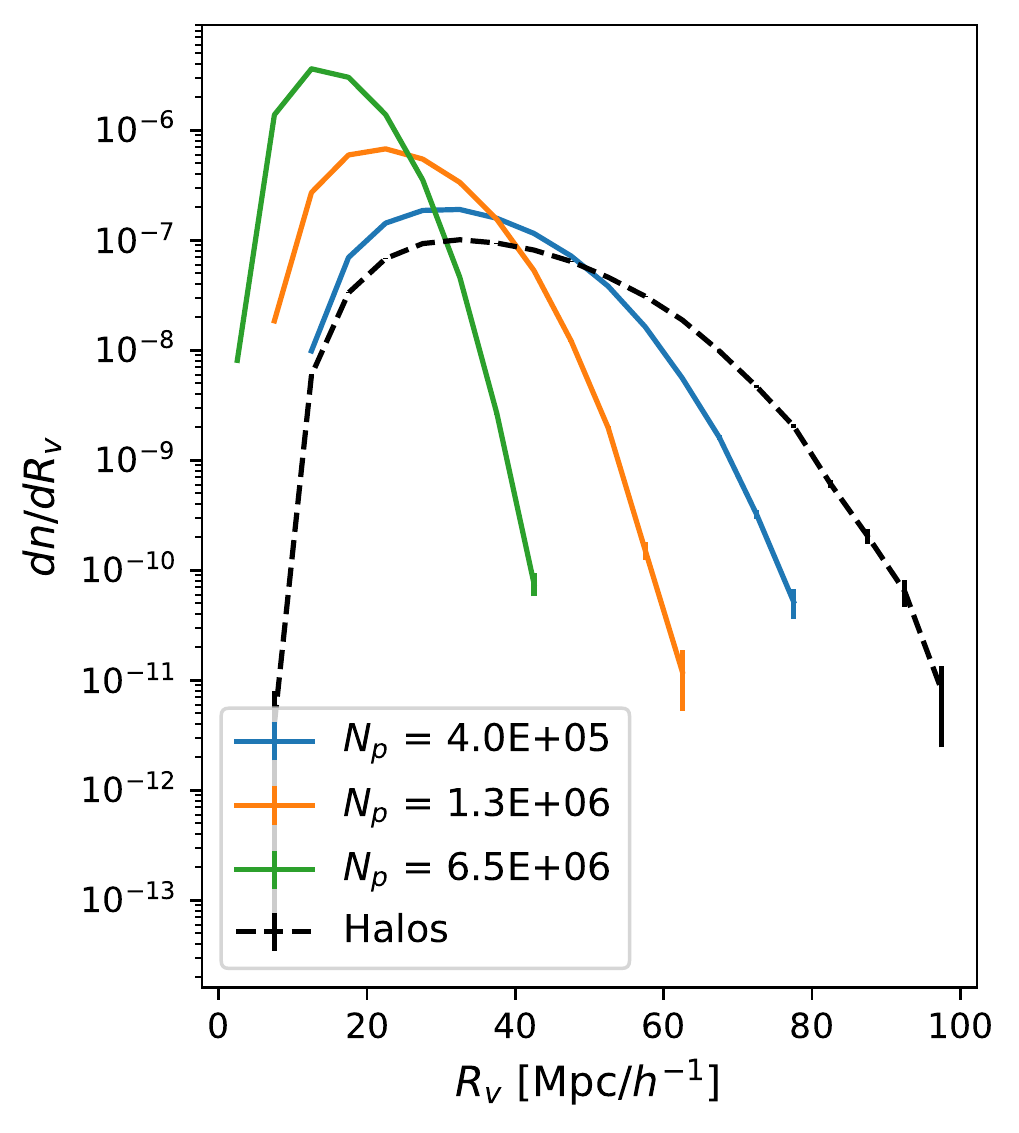}
\caption{Void size function (VSF) of spherical, Voxel and ZOBOV/VIDE voids (left to right panels) obtained from subsampled dark matter field with number of particles $N_p = 4\cdot 10^5, \, 1.3 \cdot 10^6, \, 6.5 \cdot 10^6$, and from the halo field.\label{fig:VSF_subs}}
\end{figure*}

Here we will investigate how the sparsity of a sample impacts the void centre definition and the velocity profiles around it using N-body simulations. 
The toy models in Section~\ref{sec:toy_model} demonstrated that sparsity has the potential to generate an offset between the local minimum in the full and in the subsampled matter field that is correlated with the sparse field. As a result, density and velocity profiles in the two samples can be different. We want to measure this difference in N-body simulations, and check how it depends on sparsity, i.e. on the level of subsampling performed. In what follows, all measurements made from the N-body simulations are displayed in the figures as the mean among the 50 realizations with associated errorbars computed as the standard deviation among the 50 realizations divided by $\sqrt{50}$. Errors are not displayed when plotting the ratio between measured quantities for clarity. Unless explicitly specified, in all the other cases errorbars are plotted in the figures, and if they are not visible it means that they are very small.

\subsection{Void size function}

\begin{table}
\centering
\label{tab:subs}
\caption{Number of halos and matter particles considered to build different void catalogues. For the matter field, the level of subsampling is displayed in the third column.}
\begin{tabular}{|ccc|}
\hline
Tracer & $N$ tracers & Subsampling \\ \hline
Halo & $\sim 4\cdot 10^5$ & - \\
Matter & $4\cdot 10^5$ & $0.3\%$ \\
Matter & $1.3\cdot 10^6$ & $1\%$ \\
Matter & $6.5\cdot 10^6$ & $5\%$ \\
\hline
\end{tabular}
\end{table}

First we consider voids found in the halo field and in the matter field sub-sampled to different densities. To do this, we sample the cold dark matter particles in the simulations to $N_m = 4\cdot 10^5$ (corresponding to $0.3\%$ of the total number of particles and roughly to the number of halos in the simulation), $N_m = 1.3\cdot 10^6$ ($1\%$) and $N_m = 6.5\cdot 10^6$ ($5\%$), as explained in Table~\ref{tab:subs}. Then we identify voids in these selected subsamples and in the halo field, and measure the velocity profile of the full matter field and of the selected matter particles used to identify voids around their centres. 

Figure~\ref{fig:VSF_subs} displays the size function of voids in the subsampled matter particles (colored lines) and in the halo field (black dashed lines), for different algorithms. The halo field is biased with respect to the matter field, therefore its variance is larger and it is more likely to present big fluctuations on large scales. Since big negative fluctuations give rise to voids, we expect to identify more large voids in the halo field than in the matter field. This expectation is met for all void finders considered in this paper, but to a lower extent for Voxel voids than for other void definitions. Indeed, the void size function (VSF) of Voxel voids in halos is very similar to that in the matter field subsampled to match the number density of halos.

The size function of spherical voids (left panel of Figure~\ref{fig:VSF_subs}) is relatively stable to changes in the number of dark matter particles provided a minimum tracer density threshold is reached, and it seems to be unaffected by sparsity at small void sizes. On the other hand, watershed algorithms (central and right panels) are not stable on any scales against changes in the number of tracers used to identify voids, within the ranges of tracer density tested. Modifications of the tracer density over one order of magnitude results in changes of the VSF over multiple orders of magnitude for both small and large watershed voids. For Voxel voids, some of the variation in the VSFs shown in Figure~\ref{fig:VSF_subs} may arise due to the fact that the scale of the Gaussian smoothing applied to the density field before void-finding is chosen on the basis of the tracer density, $n_t^{-1/3}$. However, the qualitative similarity of the VSFs for the Voxel and ZOBOV cases suggests that the majority of this variation is actually a function of the watershed algorithm itself, and not the smoothing scale (as ZOBOV does not use a smoothing).

\subsection{Velocity and density profiles}
\begin{figure*}
\includegraphics[width=1\textwidth]{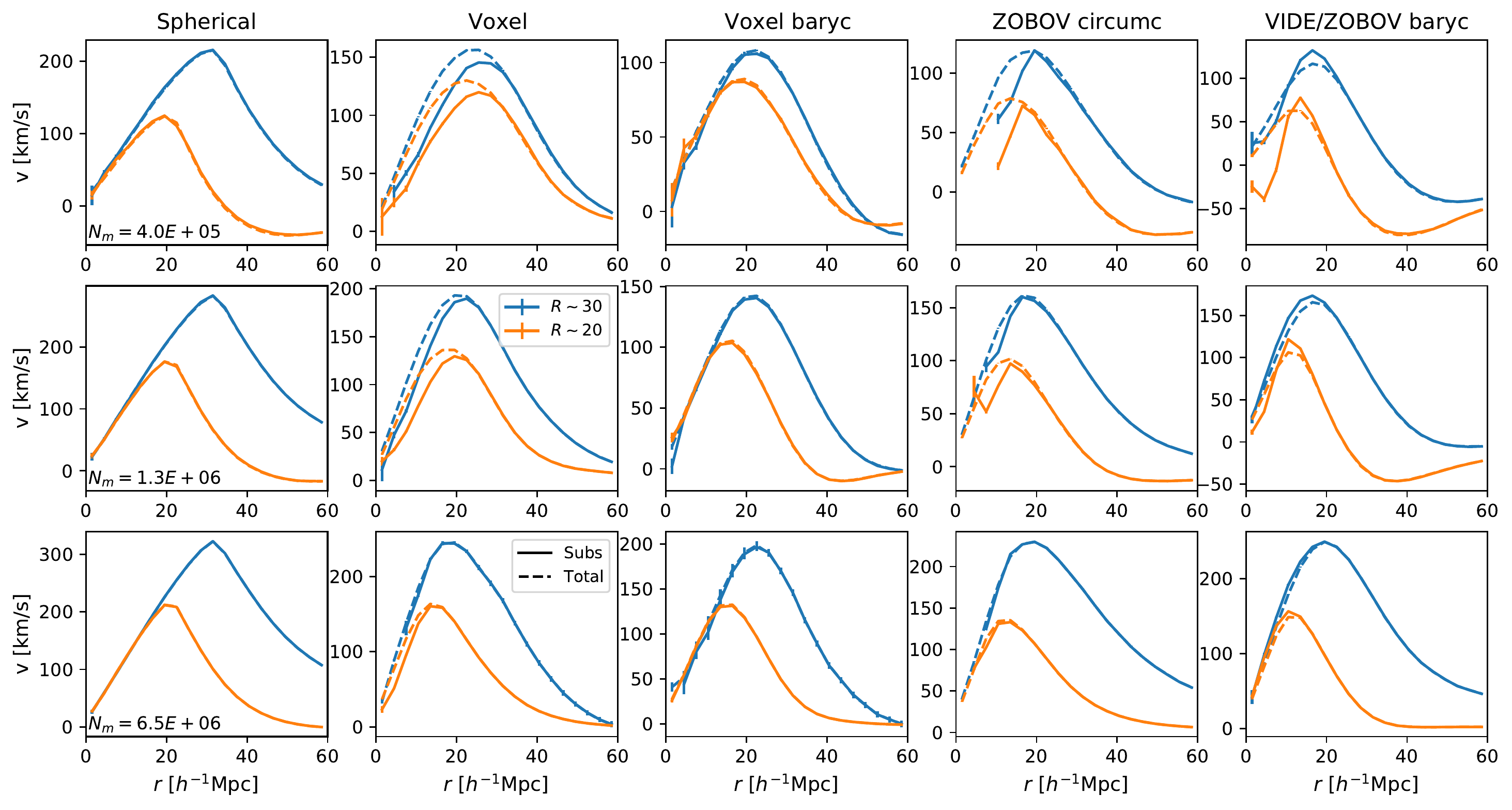}
\caption{\label{fig:vel_sub_m} Velocity profiles recovered from voids identified using the subsampled matter field. Rows show different values for the number of matter particles in the subsampled field, and columns show different void finders and centre definitions. Different colours are for different void sizes, while dashed and solid lines are for the measured profile in the total and subsampled matter.}
\end{figure*} 

\begin{figure*}
\centering
\includegraphics[width=0.99\textwidth]{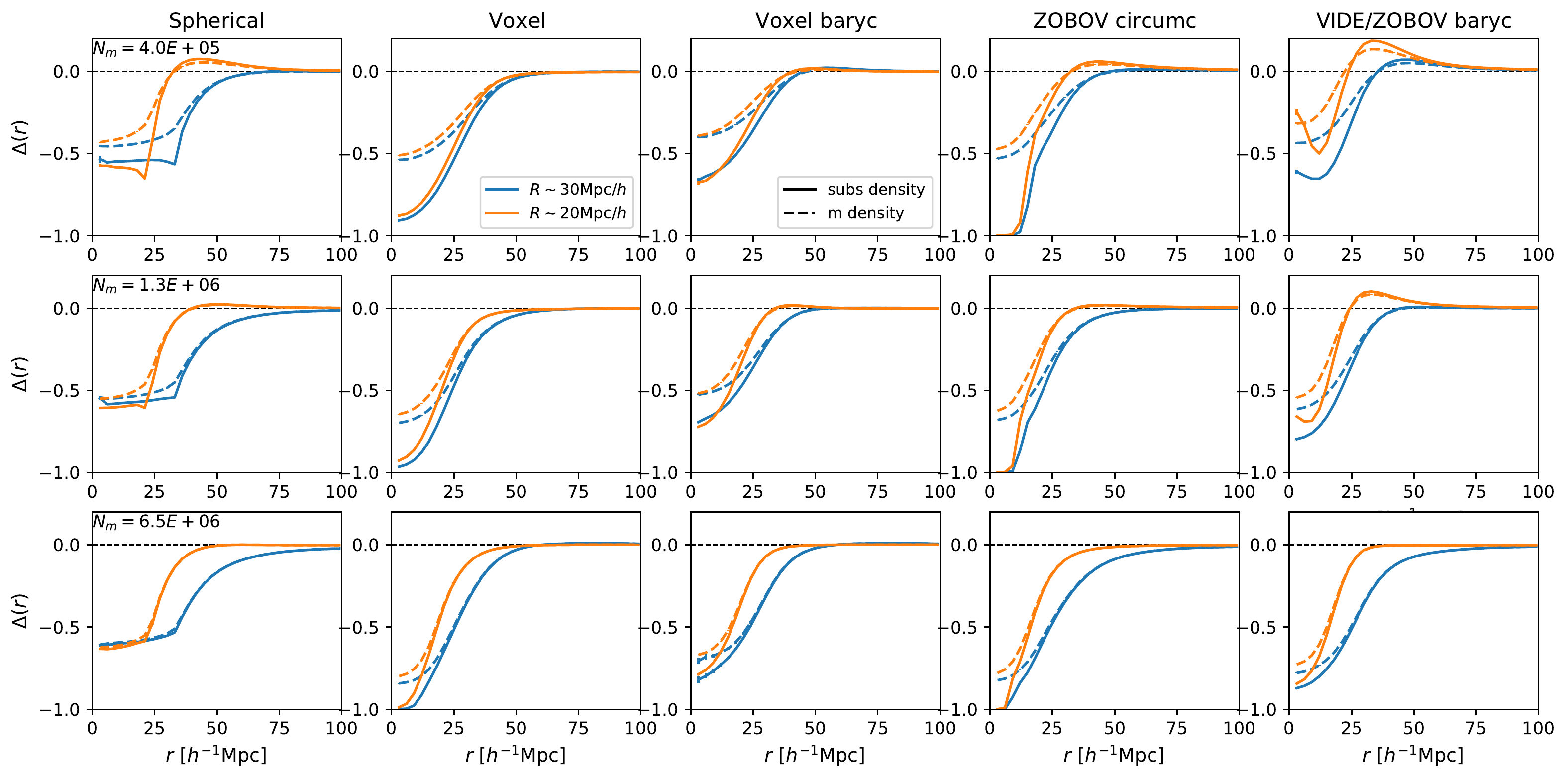}
\caption{\label{fig:delta_vm_subs} Enclosed density profiles of the subsampled (solid) and total matter (dashed) around voids identified in the subsampled field. Different colors correspond to different void sizes.}
\end{figure*}

We measure the velocity and density profiles around voids identified in samples with different levels of subsampling of the matter field. Measurements are performed around voids of similar size --- grouped in radius bins of width $5h^{-1}$Mpc --- for watershed algorithms, and voids obtained with the same smoothing scale $R_{\rm v}$ for the spherical finder. Figure~\ref{fig:vel_sub_m} shows the velocity profiles of the full matter field (dashed lines) and the subsampled matter particles used to identify the voids (solid lines). Each column displays results for a particular void finder and each row shows different level of matter subsampling. Different colors indicate different void sizes; since watershed algorithms do not identify voids within the same size range when changing the number density of tracers (see Figure~\ref{fig:VSF_subs}), we consider the largest two void sizes that are present across all the void catalogs: $R_{v}\sim 20,30h^{-1}$Mpc. Spherical voids with density threshold $\Delta_v=-0.5$ (first column) reveal matching velocities for all the subsamplings considered.  On the other hand, when the number of particles used to identify voids matches the number of halos in the simulations (first row), Voxel and ZOBOV voids exhibit differences between velocity profiles of the subsampled and full matter field. 
The velocity profile of the sparse sample used to identify voids is smaller than that of the full matter field for Voxel and ZOBOV-circumcentre voids, while the opposite happens for VIDE/ZOBOV barycentre voids. Increasing the number of matter particles used to identify voids (second row) alleviates the mismatch, which is almost absent when $N_m = 6.5\cdot 10^6$ (third row). This indicates that the sparsity of the sample can induce a discrepancy between the velocity profile of that subsample and its full realization, and this difference increases with the level of subsampling. 

Similar to our toy model in Section~\ref{sec:toy_model}, we see that the shot-noise of tracers can affect the void centre position, and in particular that the sparsity of a subample can shift the void centre location from its true position in the full sample.  
Centres that are defined locally should be affected more by sparsity than centres that are defined on a larger scale. We can check this by comparing results from the same void finder but different centre definitions, e.g. Voxel voids with the default centre definition and Voxel voids with inverse-density-weighted barycentre denotation. The barycentre definition is less local than the default one, since it takes into account the position of all particles belonging to the void --- albeit with a higher weighting attached to those in the low-density void central regions. These two cases correspond to columns 2 and 3 in Figure~\ref{fig:vel_sub_m}, respectively. The velocity profiles of full and subsampled matter around Voxel barycentres are overlapping even when considering the subsample with $N_m = 4\cdot 10^5$ (first row), while the mismatch is present in this case for standard Voxel centres. Additionally, spherical void finders have a non-local centre definition, since centres are computed by smoothing the field with a top hat filter at the void radius scale. This is likely the reason why spherical voids do not exhibit any difference between velocity profiles, even when the level of subsampling of the matter particles is very high (first row). This logic is less clear when comparing ZOBOV-circumcentres and VIDE/ZOBOV-barycentres voids (column 4 and 5 in Figure~\ref{fig:vel_sub_m}). In this case the mismatch in velocity profiles is present also for the barycentre definition when $N_m = 4\cdot 10^5$, but the velocity of the sparse sample is larger than the one of the full sample.  

Figure~\ref{fig:delta_vm_subs} shows the enclosed density profiles of the full and subsampled matter fields measured around different void types (columns) selected on particles subsampled at different levels (rows). There is a difference between the two enclosed profiles around each void type when $N_m = 4\cdot 10^5$ (first row), but the difference is very small for spherical voids but larger for watershed voids. These types of mismatches indicate either that the voids identified in the subsampled field are not true voids in the full sample, or that the minima of the overdensity field in the subsample is shifted compared to the one in the full sample. Given that the velocity profiles are also different in the inner part of the voids, the second explanation is certainly playing a role in giving the differences seen in the density profiles. These differences are mitigated when voids are identified using a larger number of dark matter particles (second and third rows).

\section{Voids in the halo field}

\begin{figure}
    \centering
\includegraphics[width=0.45\textwidth]{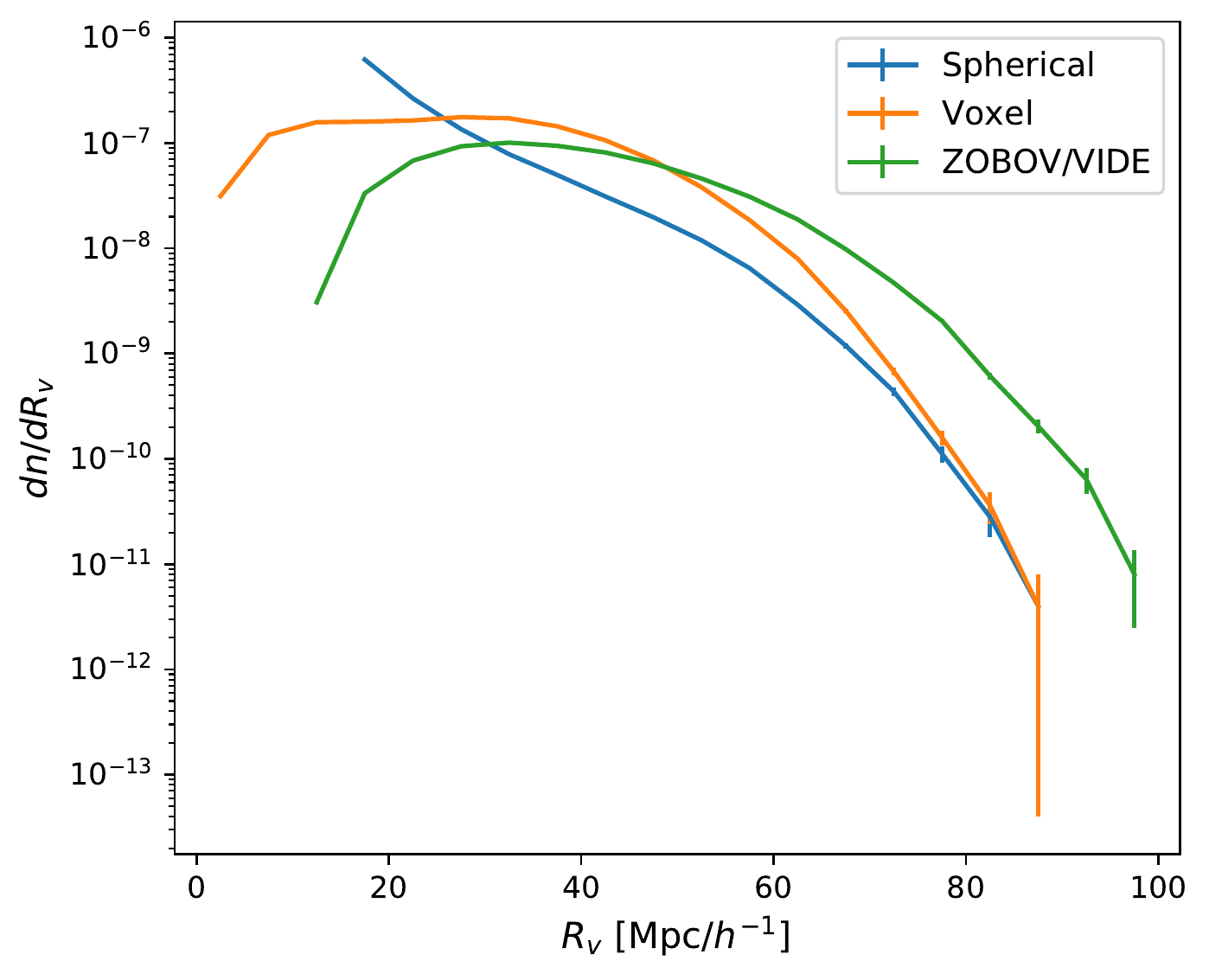}
\caption{\label{fig:VSF_h} Void size function (VSF) of voids identified in the halo field. The void finders used are spherical (blue), Voxel (orange), ZOBOV/VIDE (green).}
\end{figure}

Ultimately, we want to investigate the effect of sparsity of biased tracers as these match the situation with an observed galaxy sample. Here we consider voids identified in the halo field, and plot the void size function of spherical, Voxel and ZOBOV voids in the same panel in Figure~\ref{fig:VSF_h} (these data were previously shown in a different way in Figure~\ref{fig:VSF_subs}). The number density of voids as a function of their size is similar across the different void definitions, with ZOBOV voids tending to be larger than Voxel and spherical voids. Moreover, the number of small spherical voids increase with decreasing void size, while the number of small watershed voids is constant and decreasing with decreasing void radius.

\begin{figure*}
    \centering
\includegraphics[width=0.99\textwidth]{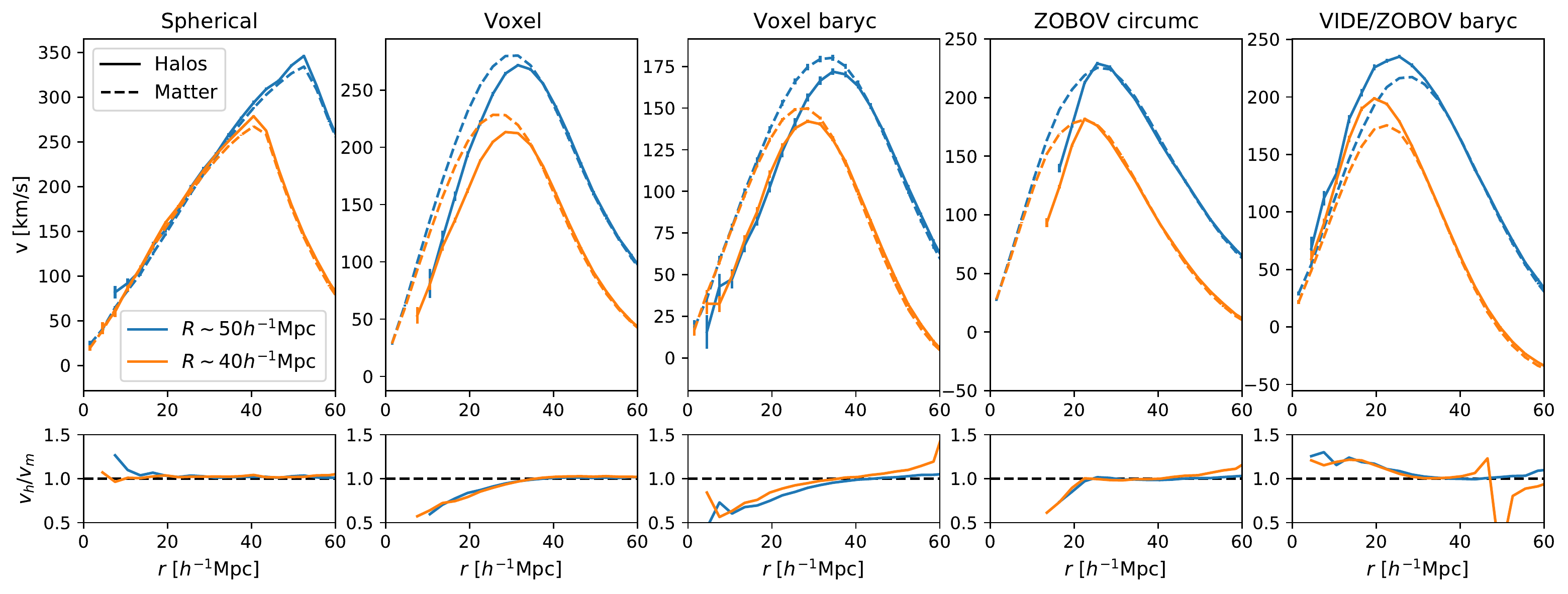}
\caption{Upper panels: Radial velocity profile of halos (solid) and matter (dashed) around voids identified in the halo field. Bottom panels: ratio between the velocity profile of halos and matter (here errorbars are not displayed). Each color corresponds to different void sizes.  \label{fig:vel_h_m}}
\end{figure*}

\subsection{Matter and Halo velocity profile}
 
Figure~\ref{fig:vel_h_m} shows the velocity profiles of halos (solid lines) and matter particles (dashed lines) around voids in the halo field. Each plot considers a different void finder and different colors report velocity profiles around voids of different sizes. The first plot from the left shows velocities around spherical voids obtained with density threshold $\Delta_v = -0.5$. In this case the matter and halo velocity profiles around the same voids (same colors) are equal on all scales within few percent --- the ratio between these two velocities is displayed in the bottom panels. The same quantities around Voxel voids are displayed in the second and third plot, with the former considering the default centre and the latter using the inverse-density-weighted barycentre definition. We can notice that in both cases the halo velocities are smaller than the matter velocities at distances smaller than $30-40h^{-1}$Mpc. The discrepancy increases going towards small separation and it is maximum near the centre, where it is of the order of $40-50\%$ (see bottom panels). Moreover, velocities around the barycentre definition (third plot) exhibits smaller differences than around the default centre definition (second plot), and they decrease with void size. This finding is consistent with the results obtained for voids in the matter field in Section~\ref{sec:subs_m}, and it is in agreement with the intuition that local definition for the void centres are more affected by the shot-noise of sparse tracers. \texttt{ZOBOV}-based voids and the velocities around them are displayed in the fourth (around circumcentres) and fifth (around volume-weighted barycentre or VIDE centres) plot. Also voids identified by this watershed algorithm present different halo and matter velocity profiles in the inner part of voids. Consistently with the results for voids in the matter field, the matter velocities are larger than the halo velocities around circumcentre (as around Voxel voids), but they are smaller around volume-weighted barycentres. Notice also that overall the maximum value reached by the velocity profile differs among void finders, with spherical voids showing the largest values. This is likely due to the fact that different void finders identify different types of voids, and the definition of $R_{\rm v}$ is not universal across void finders. It is also interesting to compare the velocity profiles of Voxel and Voxel barycentre voids, where the selected voids are exactly the same but the centre definition is different: barycenters exhibit a low velocity profile than standard centers. On the other hand, different centre definitions for ZOBOV voids do not alter the maximum of the velocity profiles, but change its minimum on large scales: velocities around VIDE/ZOBOV-barycentre voids reach lower (negative) values on large scales than ZOBOV-circumcentre voids.

\subsection{Enclosed density profile}

\begin{figure*}
    \centering
\includegraphics[width=0.99\textwidth]
{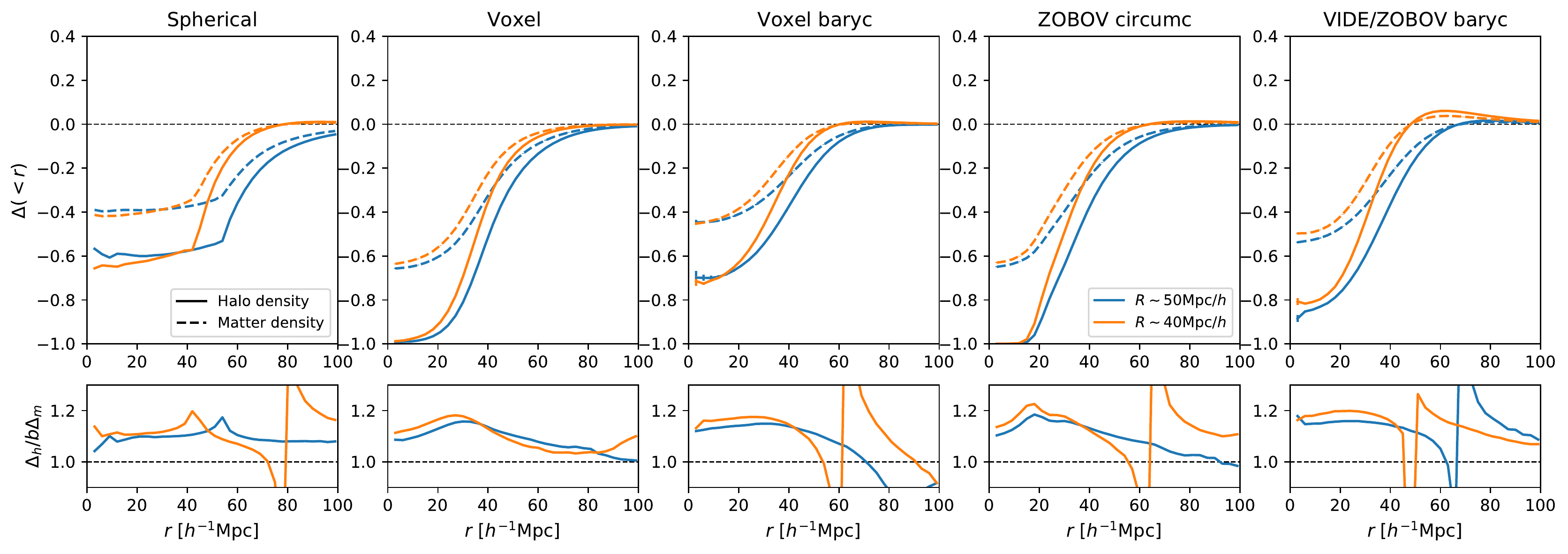}
\caption{\label{fig:delta_vh} The enclosed density profile of halos (solid) and matter (dashed) around voids identified in the halo field. Different colors correspond to different void sizes. The bottom panels show the ratio between the halo density profile and the matter density profile multiplied by the large scale halo bias. The halo bias $b=1.4$ has been computed from the ratio $P_{hm}(k)/P_{mm}(k)$.}
\end{figure*}

The discrepancy in velocity profiles around watershed voids is present at small separations $r$, corresponding to shells that are almost devoid of halos, as shown in Figure~\ref{fig:delta_vh}. Different panels refer to different void finders and show the enclosed density profile of halos (solid lines) and matter (dashed lines) around different void sizes (color coded). While the halo density around Voxel and ZOBOV-circumcentre voids reaches the value $\Delta=-1$ in the inner part of those voids, the analogous density around spherical voids and barycentres is always larger than $\sim -0.8$. 

\begin{figure*}
    \centering
\includegraphics[width=0.99\textwidth]
{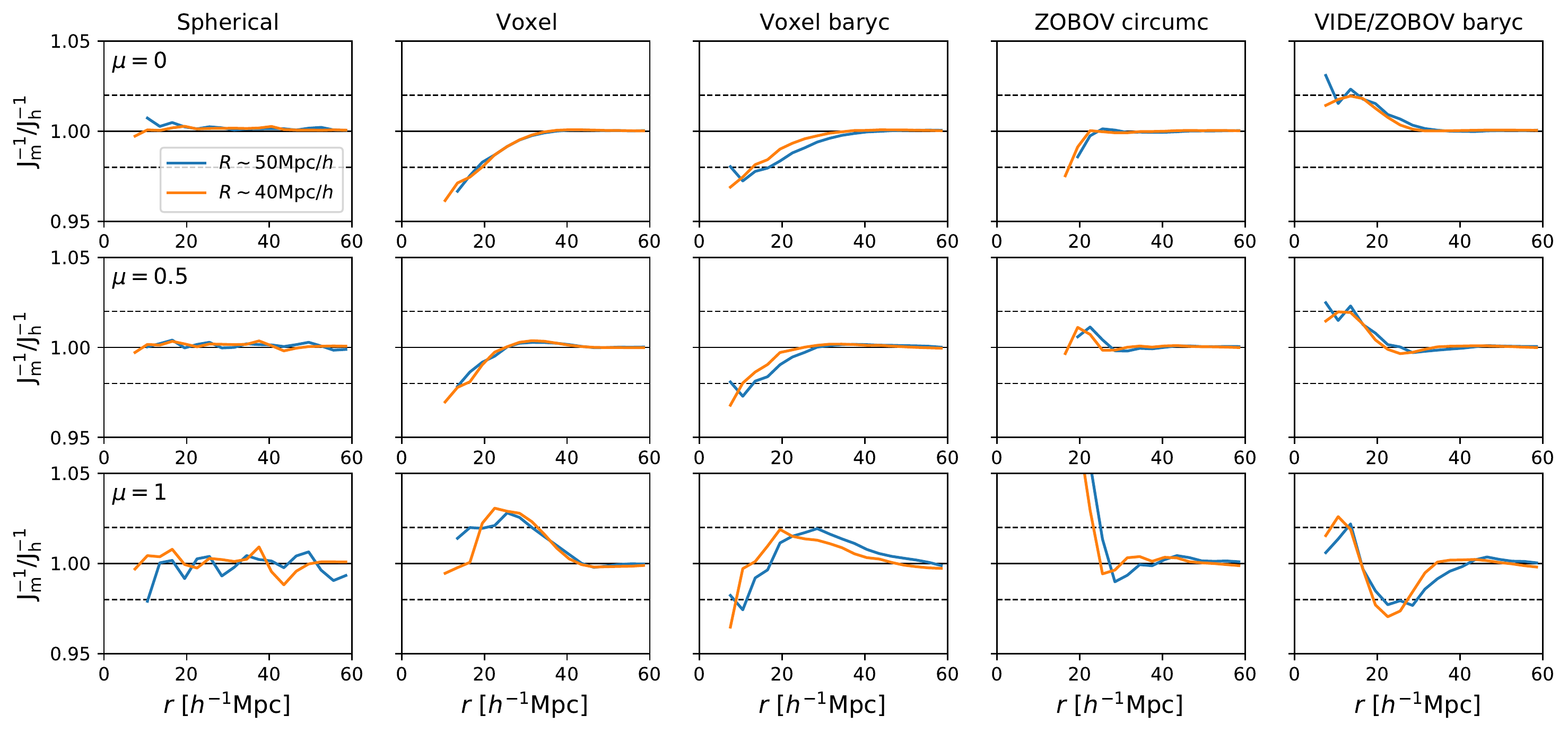}
\caption{Relative difference between the inverse of the Jacobian of the transformation of coordinates from real to redshift space obtained using the matter velocity and the halo velocity; different colors refer to  different void sizes. The upper/central/lower panels show results for $\mu=0/0.5/1$. \label{fig:jacobian_h_m}}
\end{figure*}
Halos are biased tracers of the matter field, and the population considered here has large scale bias larger than one. It is therefore expected that the matter density profiles are closer to the mean density than the halo profiles. The bottom panels of Figure~\ref{fig:delta_vh} show the ratio between the enclosed halo profile and the matter profile multiplied by the large scale linear halo bias measured from the halo-matter cross-power spectrum in simulations, $\Delta_h/b\Delta_m$. The ratio is not equal to $1$ at large distances $r$, as one would naively expect.  This ratio is approximately scale independent around the largest spherical voids, and presents a more pronounced scale dependency around watershed voids, and it is almost scale independent on small scales around barycentre voids (spikes are always present where the matter profile $\Delta_m$ crosses zero). The fact that this ratio is different from $1$ on large and small scales indicates that the halo bias around voids is different from the halo bias of the full halo catalog. \cite{Pollina_2019} found the same result inspecting the galaxy and cluster bias in VIDE voids. Our analysis suggests that this is due to the sparsity of halos and that it depends on the void centre definition in different void finders.

\subsection{Jacobian for redshift space analysis}

The Jacobian of the real-space to redshift-space transformation used in Equation~\ref{eq:xi_r_to_s} is \citep{Cai_2016}
\begin{equation}
    J = 1+(1-\mu^2) \frac{v_r}{aHr} + \frac{\mu^2}{aH}\frac{\partial v_r}{\partial r},
    \label{eq:J}
\end{equation}  
and it depends on the galaxy velocity profile and its spatial derivative. In the literature it is often assumed that the galaxy and the matter velocity profiles match on all scales. Then the matter velocity profile is used to model the galaxy motion around voids via the continuity equation that maps the matter density profile to the matter velocity profile. However, if the matter and galaxy velocity profiles do not match at small separations, as shown in Figure~\ref{fig:vel_h_m} where halos are used to mimic the galaxy behaviour, then assuming that they are equal can introduce a systematic error in the estimation of the Jacobian in Equation~\ref{eq:J}. We quantify this systematic error in Figure~\ref{fig:jacobian_h_m}, where each plot considers the relative difference between the inverse of the Jacobian using the matter versus halo velocity profile in Equation~\ref{eq:J} evaluated at different values of $\mu$, and each panel reports the results for a different void finder. When the ratio is not equal to $1$, the two Jacobians are not identical. The black dashed lines indicate where the ratio of the inverse Jacobians differs from unity by $\pm2\%$. 
The difference in matter and halo velocity profiles around watershed voids translates into a difference of relative Jacobians of up to about $5\%$ in the inner part of voids, reducing to larger scales. The offsets on larger scales are larger along the line-of-sight direction (i.e., at $\mu \simeq 1$). In the subsequent figures we have omitted the spherical void finder, since Figures~\ref{fig:vel_h_m} and~\ref{fig:jacobian_h_m} showed that the velocity profiles and Jacobians of the matter and haloes match very well for that case.

\subsection{Consequences for $f\sigma_8$}

\begin{figure*}
\centering
\includegraphics[width=0.99\textwidth]{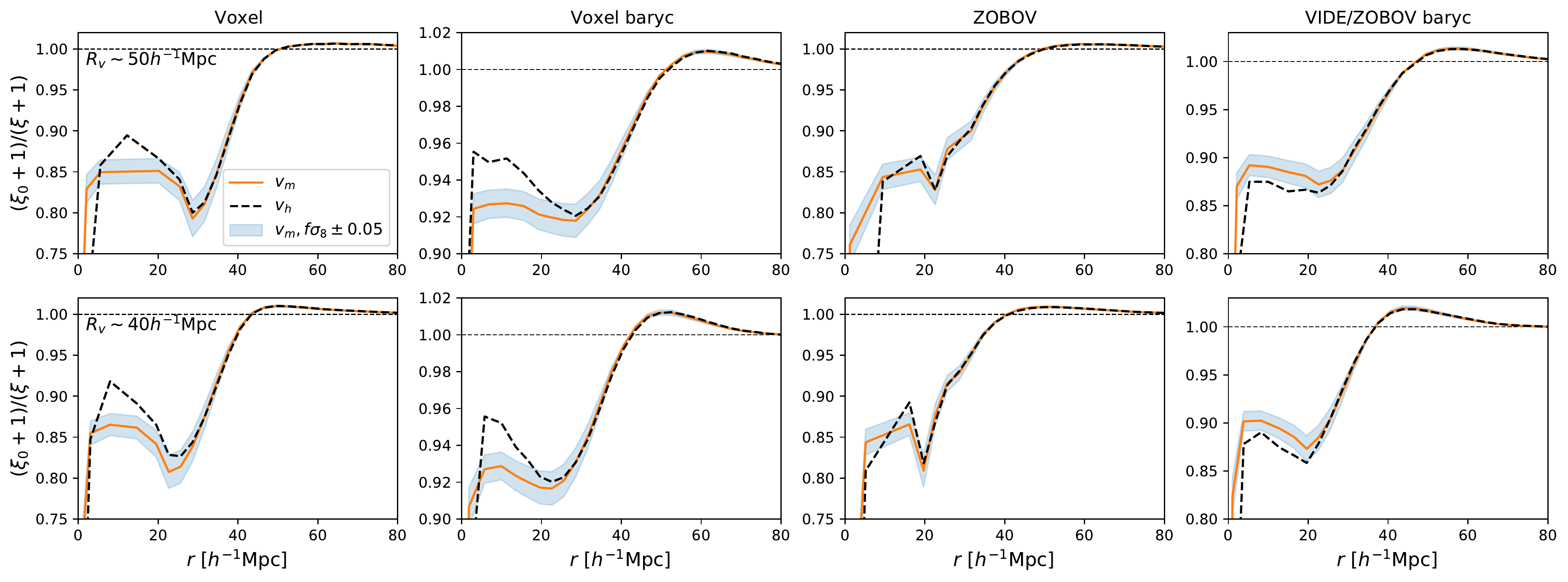}
\caption{Ratio of the monopoles of the void-halo correlation function in redshift space and the real space. The redshift space part has been computed using Eq.~\ref{eq:xi_r_to_s} with the real space counterpart measured in simulations. Each column shows results for a different watershed void finder and voids have been selected in void size bins: $R_{\rm v}=50-55~h^{-1}$Mpc (upper row) and $R_{\rm v}=40-45~h^{-1}$Mpc (lower row). The Jacobian in Eq.~\ref{eq:xi_r_to_s} has been calculated using the halo velocity profile (orange lines), matter velocity profile (black dashed lines). To roughly estimate the error introduced by using the matter velocity, rather than the halo velocity, we calculated the Jacobian using the matter velocity re-scaled by a wrong value for $f \sigma_8$ (green and blue lines) corresponding to a variation indicative of the statistical uncertainty in measurement of $f\sigma_8$ from current void-galaxy cross-correlation analyses \citep{Nadathur_2019c, Woodfinden_2022}.}
\label{fig:x0s}
\end{figure*}

\begin{figure*}
\centering
\includegraphics[width=0.99\textwidth]{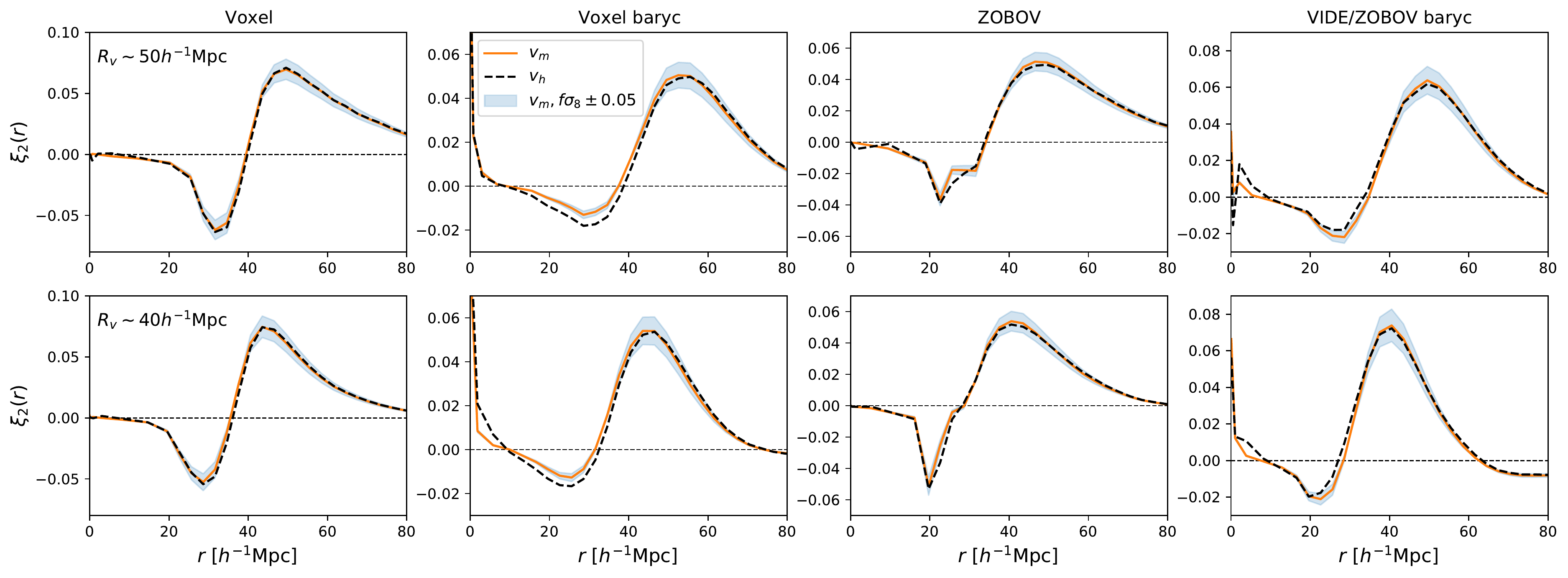}
\caption{Quadrupole of the void-halo correlation function computed using Eq.~\ref{eq:xi_r_to_s}, using the real space correlation measured in simulations. Panels and colors are the same as in Figure~\ref{fig:x0s}}
\label{fig:x2s}
\end{figure*}

Fitting to the redshift-space void-galaxy correlation function can inform us about the growth of cosmological structure by constraining the combination of parameters $f\sigma_8$, where $f$ is the linear growth rate and $\sigma_8$ is the amplitude of matter perturbations at redshift $z=0$ (e.g. \citealt{Woodfinden_2022}). In the literature, the galaxy velocity profile that enters in the Jacobian (Equation~\ref{eq:J}) is assumed to match the corresponding matter velocity profile and is usually expressed in terms of the enclosed matter density profile via Equation~\ref{eq:d_to_v}. Since both $f$ and the enclosed density profile in Equation~\ref{eq:d_to_v} are evaluated at a fiducial cosmology, the resulting velocity $v_{\rm fid}(r)$ is re-scaled accordingly to
\begin{equation}
    v(r) = \frac{f\sigma_8}{\left[f\sigma_{8}\right]_{\rm fid}} \, v_{\rm fid}(r),
\end{equation}
to obtain an estimation of $f \sigma_8$.

Since the two velocity profiles can differ, we predict the impact of using the matter velocity in the void-galaxy correlation function analysis in place of the correct (galaxy) velocity, and in particular its consequences on the estimation of $f\sigma_8$. A quantitative analysis would involve running a full parameter estimation pipeline and would need to take into account the different ways to model for the void-galaxy multipoles available in the literature. Therefore here we only provide a qualitative comparison of differences to the predicted multipoles of the correlation function that arise from (a) using the matter velocity profile instead of the galaxy one, and (b) using a fixed velocity profile but with shifted values of $f\sigma_8$.

First, we measure the real space void-halo correlation function in the Quijote simulations, then we map it to redshift space via Equation~\ref{eq:xi_r_to_s} using different velocity profiles in the Jacobian, then we evaluate the multipoles as
\begin{equation}
    \xi_l(s) = \frac{2l+1}{2}\int_{-1}^1 d\mu \, L_l(\mu) \, \xi(s,\mu),
\end{equation}
where $L_{0,2}(\mu) = 1,\, (3\mu^2-1)/2$ are the Legendre polynomials used to obtain the monopole and quadrupole of the correlation function, respectively. The results are shown in Figures~\ref{fig:x0s} and~\ref{fig:x2s}, with the former  displaying the ratio between the redshift space monopole calculated in distinct ways and the measured real space correlation, and the latter reporting the model quadrupoles. Results for different void finders are shown in different columns, and rows indicate different void sizes. Multipoles computed using the (correct) halo velocity profile are denoted in black dashed lines, and those using the matter velocity profile in orange. We consider a re-scaling of the matter velocity profile by values of $f\sigma_8$ that are $1\sigma$ away from the true value in the simulations, with $\sigma=\pm 0.05$ being a value indicative of the statistical uncertainty in measurement of $f\sigma_8$ from current void-galaxy cross-correlation analyses \citep{Nadathur_2019c, Woodfinden_2022}. 

Changes in the monopole with modifications in $f\sigma_8$ (compare orange, blue and green lines) are visible only on scales smaller than $\sim40~h^{-1}$Mpc. These are also the scales where the monopole obtained from the matter velocity profile differs from the one obtained using the halo velocity profile. The differences between using $v_h$ or $v_m$ are comparable to, or larger than, the difference induced by a $1\sigma$ shift in $f\sigma_8$, and are present in all watershed void finders. The discrepancy is particularly visible on scales smaller than $30~h^{-1}$Mpc, where it is larger than the one induced by the $1\sigma$ shift in $f\sigma_8$ for Voxel and Voxel barycentre voids, it is of the order of $1\sigma$ variation for VIDE/ZOBOV barycentre voids and smaller than that for ZOBOV-circumcentre voids --- although these voids are almost empty on those scales and the effect seems to be noise dominated.

The quadrupoles exhibit changes up to larger scales --- up to $r<70-90~h^{-1}$Mpc depending on the void size and the void finder considered --- when varying $f\sigma_8$. On the other hand, the differences caused by using the matter or the halo velocity profile show up over a limited range of scales smaller than $\sim30 ~h^{-1}$Mpc. Indeed for Voxel voids (left column of Figure~\ref{fig:x2s}) the effect of switching between $v_m$ or $v_h$ appears to be negligible, but for other void types differences can be more substantial.

Overall, we conclude that systematic changes in the predicted void-galaxy cross-correlation multipoles from using models that describe the matter velocity profile $v_m$ rather than the correct galaxy/halo velocity profile $v_h$ can be substantial, particularly in the monopole. These systematic modifications can be larger than those arising from changes in the model parameter $f\sigma_8$ over the range of current observational uncertainty, especially at scales $r\lesssim30-40\,h^{-1}$Mpc. However, these modifications do not have the same scale-dependence as those arising from changes in $f\sigma_8$. Thus it is possible that they can be distinguished, and would not lead to large systematic biases in the inferred $f\sigma_8$ values recovered from fits to observations. Indeed, careful analyses of mocks suggests that the systematic error remains a small fraction of the total error budget for current analyses \citep[e.g., see][]{Woodfinden_2022}, but this conclusion may no longer hold for future analyses of data from DESI\footnote{\url{https://www.desi.lbl.gov}} and Euclid\footnote{\url{https://www.euclid-ec.org}} with much higher statistical precision. In the mean time, it is clear that at the very least neglecting this effect could lead to a worse model fit to the data. 

\section{Conclusions}

Upcoming spectroscopic surveys will observe the spectra of a large number density of galaxies in a huge volume, and they will draw a very detailed map of the Universe, including the less populated regions such as cosmic voids. Voids have been shown to be useful for studying cosmology, and many void-related statistics have been proposed and considered in the literature. Since voids are identified in the galaxy field using galaxies, which are not only biased but also sparse tracers of the matter field, there might be some concerns regarding the impact of galaxy sparsity on void identification and void-related observables. 

In this work we have investigated the impact of sparsity on the galaxy density and velocity profiles measured around voids. Using toy models, we have shown how sparsity can affect them, through the correlation of the void centre with the sample from which the void is found, which is also the sample used to define the profiles. In particular, we showed that a highly sub-sampled matter field can lead to void centres that are offset with respect to the void centres in the full matter field. This shift is caused by shot-noise in the sparse sample and leads to void centres correlated with the noise in the sparse sample. As a consequence, the density and velocity profiles of the full and sub-sample around the same shifted centres are different in the inner part of the voids. While these effects are caused by the sparsity of the tracers used to identified voids, their extent depends on the void and void-centre definitions.   

We have quantified these effects using a subset of the Quijote simulations using three void finders (spherical, Voxel and \texttt{ZOBOV}) and different void-centre definitions that generated five different types of void catalogs. First, we investigated sparsity in the absence of bias, using the matter field randomly subsampled to different numbers of dark matter particles, and identified voids in them. The velocity profiles of the full and subsampled matter field match on all scales when computed around spherical voids. Differences are present when watershed voids are considered, but they are mitigated when considering a larger number of tracers in the subsampled field. Moreover, discrepancies seen in Voxel voids around the default centres are erased when considering inverse-density-weighted barycentres, and similarly the differences around ZOBOV-circumcentres appear to be somewhat mitigated when introducing volume-weighted barycentres (VIDE voids). This result is expected, since barycentres are a non-local definition for the void centre position, and thus should be less affected by shot-noise that becomes relevant in the inner part of voids where very few galaxies can be found.

After that, we performed a similar analysis around voids identified in the halo field. Halos are both sparse and biased tracers of the matter field, and are therefore used to describe and mimic the effect of sparsity in galaxies. Our results showed that the halo and matter velocity profiles around spherical voids are matching on all scales, but they differ around all the other watershed void and void-centre definitions considered. This difference increases with decreasing distance from the void centre, and can reach discrepancies of around $40\%$ in the the very inner part of voids. 

Velocity profiles are commonly used to model the Jacobian of the real to redshift space mapping in void-galaxy cross-correlation analysis. Often, the continuity equation is used to map the enclosed matter density profile to the matter velocity profiles, which is then used to describe the galaxy velocity profile assuming that they are identical. We quantified the systematic error introduced by this assumption in the Jacobian. Differences in the matter and halo velocity profiles of the order of $20-40\%$ seen in the simulation data translate to errors of $\sim5\%$ in the Jacobian. We then calculated the monopole and quadrupole of the void-halo correlation function starting from the measured real space counterpart and using distinct Jacobians obtained from the matter and the halo velocity profiles, and matter velocity profiles re-scaled to different values of $f\sigma_8$. We consider values of $f\sigma_8$ that are varied from the value of the simulations by the uncertainties of the RSD analysis with the void-galaxy cross-correlation in \cite{Nadathur_2019c} and \cite{Woodfinden_2022}, that considered multiple redshift bins between $z=0.07$ and $z=1$ in the Sloan Digital Sky Survey \citep{2000AJ....120.1579Y}. The discrepancy on small scales between monopoles computed using the matter and halo velocity profiles are comparable to, or larger than, the change introduced by the re-scaling of the matter velocity profile with the considered values of $f\sigma_8$. This discrepancy are present for all watershed void finders and all center definitions considered. Similar, but less pronounced differences are present in the quadrupoles. While a simple re-scaling using a different value of $f\sigma_8$ cannot account for the scale-dependent modification of the multiples induced by the use of the matter velocity profile rather than the (correct) halo velocity profile, the forecast statistical uncertainties in $f\sigma_8$ in future data \citep{Hamaus_2022} will also be much smaller than the current limits. This indicates that future RSD analysis with void-galaxy correlation function need to take into account the effect of sparsity on the halo/galaxy velocity profiles to avoid a biased estimation of $f\sigma_8$ when using watershed algorithms. Spherical voids appear to be much less (or almost not) affected by sparsity in terms of both void size function and velocity profiles. This indicates that spherical void finders may be more robust than watershed algorithms, and should encourage their implementation in future analysis with cosmic voids.

\section*{Acknowledgements}

Research at Perimeter Institute is supported in part by the Government of Canada through the Department of Innovation, Science and Economic Development Canada and by the Province of Ontario through the Ministry of Colleges and Universities. This research was enabled in part by support provided by Compute Ontario (computeontario.ca) and the Digital Research Alliance of Canada (alliancecan.ca). SN is supported by an STFC Ernest Rutherford Fellowship, grant reference ST/T005009/2. 

For the purpose of open access, the authors have applied a CC BY public copyright licence to any Author Accepted Manuscript version arising.

\section*{Data Availability}

Data supporting this research are available on reasonable request from the corresponding author.



\bibliographystyle{mnras}
\bibliography{example} 




\appendix

\section{Volume-weighted velocity profile}

\begin{figure*}
    \centering
\includegraphics[width=0.8\textwidth]{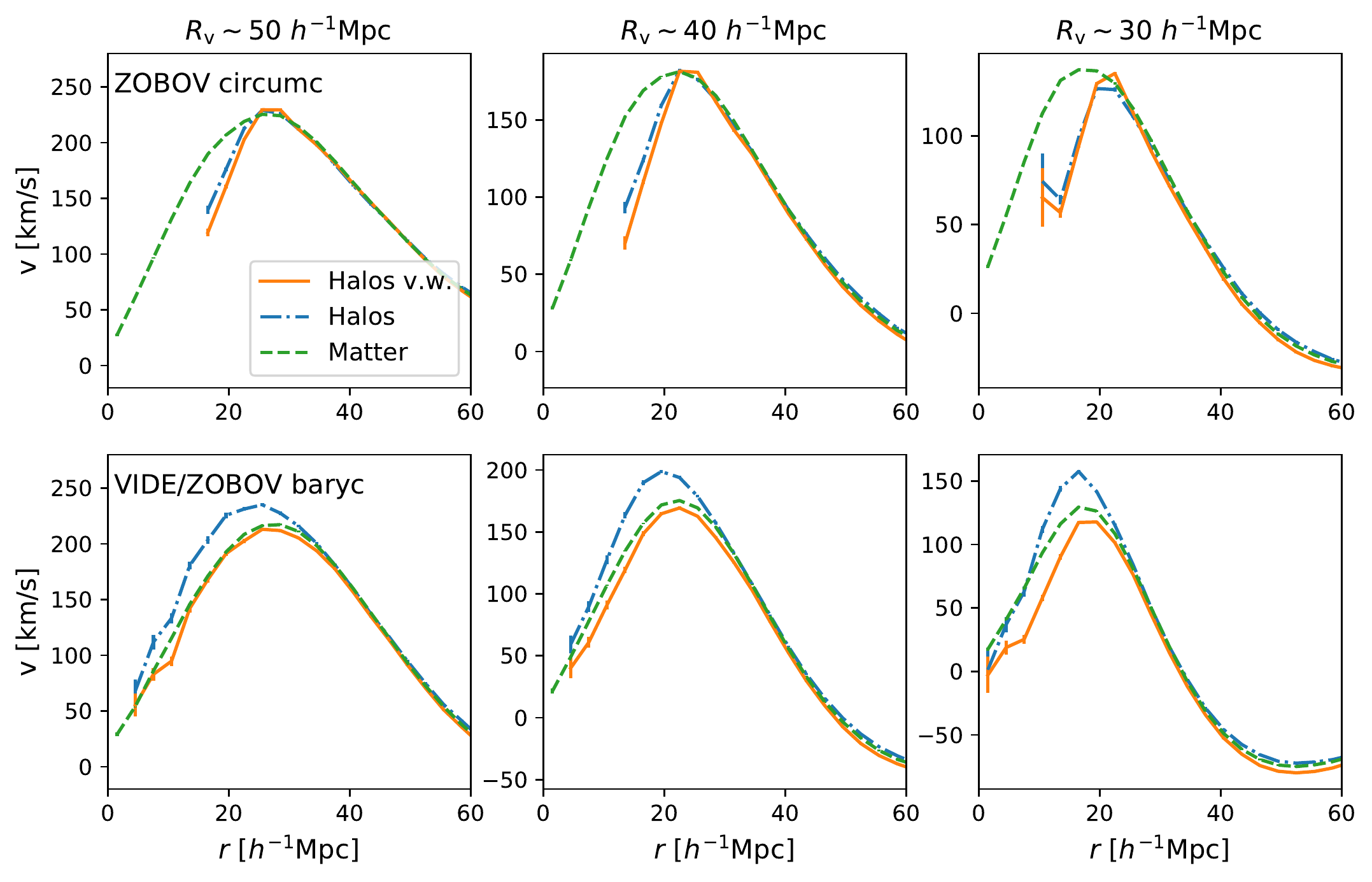}
\caption{\label{fig:vw_vel} Velocity profile around ZOBOV circumcentre (top) and VIDE/ZOBOV barycentre (bottom) voids identified in the halo field and with size $R_{\rm v} = 50,40,30~h^{-1}$Mpc (left to right columns). Volume-weighted halo velocity profiles are displayed in orange, while standard halo and matter velocity profiles are indicated as blue dash-dotted and dashed green lines, respectively.}
\end{figure*}

A different estimator for the velocity profile around watershed voids has been proposed in the literature \citep{Hamaus_2014}, the volume-weighted velocity profile
\begin{equation}
    v_V(r) = \frac{\Sigma_{i=1}^{N_v} \Sigma_{j=1}^{N_p} \, I_r(r_{ij})\, \vec{v}_j \cdot \hat{r}_{ij} \, V_c^j}  {\Sigma_{i=1}^{N_v} \Sigma_{j=1}^{N_p} \,  I_r(r_{ij})\, V_c^j}\, ,
\end{equation}
which is similar to Equation~\ref{eq:vel} except for the addition of $V_c^j$ that is the volume of the Voronoi cell around the object (galaxy) $j$. This estimator weights each object by the volume of its Voronoi cell, that is a proxy for the inverse of the local density around it. In this way the volume-weighted velocity profile accounts for the fact that low density regions are traced by very few objects, and thus tries to correct for effects coming from the sparsity of the tracers. We will test if it can estimate a tracer velocity profile that match the matter velocity profile.

We measure the volume-weighted velocity profile of halos around ZOBOV voids identified in the halo field in the simulations, and report the results in Figure~\ref{fig:vw_vel} together with the standard velocity profile of halos and matter computed as in Equation~\ref{eq:vel} and already shown in Figure~\ref{fig:vel_h_m}. When considering ZOBOV-circumcentre voids (top panels), the volume-weighted and the standard velocity profile of halos overlap around large voids; the results are different for VIDE/ZOBOV barycentre finders (bottom panels). In that case, the volume-weighted velocity profile of halos can match the matter velocity profile better than the standard estimator, but only when considering large voids. The volume-weighted estimator up-weight halos in the low halo density regions and down-weight those in denser areas. If spatial distributions of halo (subsample, orange line) and matter (full, green line) display a relative off-set as in the bottom panel of Figure~\ref{fig:toy_model_1D_x} and a velocity outflows as the top panel, then up-weighting the empties part effectively moves up the $x<0$ part of the halo distribution making it closer to the matter configuration, while down-weighting the denser regions brings down the $x>0$ part of the halo distribution getting it also closer to the matter one. It is then a way to recover the matter profile around the void centre in the sparse sample. Unfortunately this is not useful since the other way around is needed in the RSD analysis with void-galaxy correlations, i.e. we need to recover the velocity profile of the sparse tracer from a theory model that describes the matter one.


\bsp	
\label{lastpage}
\end{document}